\documentclass[a4paper,11pt]{article}
\usepackage{amsmath}
\usepackage{amsfonts}
\usepackage{amssymb}
\usepackage{mathbbol}
\usepackage{slashed}
\usepackage{graphicx}
\usepackage[dvips,a4paper,textwidth=150mm,bmargin=35mm,tmargin=35mm]{geometry}
\usepackage{braket}
\usepackage{dsfont}
\usepackage[all]{xy}

\usepackage{citesort}
\numberwithin{equation}{section}

\renewcommand\Re{\operatorname{Re}}
\newcommand\Rens{\operatorname{Re}\!}
\renewcommand\Im{\operatorname{Im}}
\newcommand\Imns{\operatorname{Im}\!}
\newcommand\iu{i}
\newcommand\ee{e}
\newcommand\mev{\,\mathrm{MeV}}
\newcommand\gev{\,\mathrm{GeV}}

\newcommand\intd{\mathrm{d}}
\newcommand\Tr{\mathrm{Tr}}

\newcommand\tr{\mathrm{tr}}

\begin{document}
\title{\bf The chiral and deconfinement crossover transitions:\\
PNJL model beyond mean field
\footnote{Work supported in part by BMBF, GSI, INFN, the DFG excellence cluster ``Origin and Structure of the Universe'' and by the Elitenetzwerk Bayern.}}
\author{
S. R\"o{\ss}ner$^{a}$, T.~Hell$^{a}$, C. Ratti$^{b}$
 and W. Weise$^{a}$\\
\\{\small $^a$ Physik-Department, Technische Universit\"at M\"unchen, D-85747 
Garching, Germany}\\
{\small $^b$ Department of Physics \& Astronomy, State University of New York,}\\
{\small Stony Brook, NY 11794-3800, USA }
}

\date{October 13, 2008}
\maketitle

\begin{abstract}
The Polyakov loop extended Nambu and Jona-Lasinio model (PNJL model) in a mean field framework shows astonishingly good agreement with lattice QCD calculations which needs to be better understood. 
The present work reports on further developments concerning both Polyakov loop and mesonic fluctuations beyond mean field approximation.
Corrections beyond mean field are of special interest for the thermal expectation values of the Polyakov loop $\braket{\Phi}$ and its conjugate $\braket{\Phi^*}$, which differ once the quark chemical potential is non-zero.
Mesonic fluctuations are also considered with emphasis on the role of pionic modes. 
\end{abstract}

\section{Introduction}

Exploring the thermodynamic properties of strongly interacting matter has become a central theme of high-energy nuclear physics in recent years. 
On the theoretical side, great progress has been achieved thanks to lattice calculations solving discretised QCD numerically.
At the present stage a major part of the numerical expense at finite quark chemical potential is caused by the fermion sign problem.
The three most promising ways to address this difficulty are multi-parameter re-weighting techniques \cite{Fodor:2001pe,Fodor:2002km}, analytic continuation from imaginary chemical potentials \cite{deForcrand:2002ci,deForcrand:2003hx} and Taylor series expansion methods \cite{Allton:2002zi,Allton:2003vx,Ejiri:2006ft,Allton:2005gk,Boyd:1996bx,Kaczmarek:2002mc,Boyd:1995cw}.

It is an important task of effective field theories and models to reveal principal mechanisms and their functioning behind the otherwise hidden mechanisms of lattice QCD. 
Alongside with other approaches \cite{Schaefer:2007pw,Barducci:1989euBarducci:1989wi,Berges:1998rc} the Nambu and Jona-Lasinio (NJL) model \cite{Ratti:2005jh,Sasaki:2006ww,Ghosh:2006qh,Mukherjee:2006hq,Zhang:2006gu,Abuki:2008nm,Fukushima:2002ew_Fukushima:2003fm,Fukushima:2003fw_Hatta:2003ga} is an approach that successfully describes spontaneous chiral symmetry breaking. 
We use the NJL model with $N_{\mathrm{f}} =2 $ quark flavours as one of our starting points.
In the NJL model gluonic degrees of freedom are ``integrated out''. 
The role of the gluons is assumed to be modelled in part by a local effective quark colour current interaction.
From this effective interaction a Fierz transformation generates various quark-antiquark and diquark coupling terms.
By integrating out the gluons the local $\mathrm{SU}(3)_\mathrm{c}$ gauge symmetry is lost.
As a consequence the NJL model, equipped only with a global $\mathrm{SU}(3)_\mathrm{c}$ symmetry, does not feature confinement.

To bring aspects of confinement back into the model an additional homogeneous temporal background field with standard $\mathrm{SU}(3)_\mathrm{c}$ gauge invariant coupling to the quarks is introduced \cite{Fukushima:2002ew_Fukushima:2003fm,Fukushima:2003fw_Hatta:2003ga}, implementing the Polyakov loop. 
This generalised NJL model, with Polyakov loop dynamics incorporated, is called PNJL model. 
In the limit of static quarks (i.\,e.\ in pure gauge QCD) the Polyakov loop serves as an order parameter for confinement.
In this limit the $\mathrm{Z}(3)$ centre-symmetry of the $\mathrm{SU}(3)_\mathrm{c}$ gauge group is unbroken, and the deconfinement transition is connected with the spontaneous breakdown of this symmetry.
In the presence of dynamical quarks the $\mathrm{Z}(3)$ centre-symmetry is broken explicitly, such that the deconfinement transition is no longer a phase transition in the strict sense. 
Nevertheless, the Polyakov loop shows a rapid crossover near the deconfinement transition, still permitting to use the Polyakov loop as a measure for deconfinement. 
The confining gluon dynamics that was lost in the NJL model is now re-introduced via an effective potential.
This potential is part of a Ginzburg-Landau model for confinement in the static quark limit. 
The information necessary in order to specify the effective potential is extracted from pure glue lattice QCD calculations \cite{Boyd:1995cw}. 

In Sec.~\ref{sec:pnjlmodel} we review the PNJL model  \cite{Ratti:2005jh,Sasaki:2006ww,Ghosh:2006qh,Mukherjee:2006hq,Zhang:2006gu} at the level of the mean field approximation. 
When implementing the Polyakov loop extension to the NJL model, it is important to pay special attention to the fermion sign problem. 
As the Polyakov loop is coupled to the NJL model in analogy to QCD via minimal substitution the fermion sign problem in QCD and in the PNJL model appears on equal footing.
Issues arising from the fermion sign problem in the PNJL model are discussed in Sec.~\ref{sec:corr}. 
In the present paper these issues are addressed more explicitly than in previous work \cite{Ratti:2005jh,Roessner:2006xn,Ratti:2006wg}. 
The method developed in this improved treatment introduces a systematic expansion around a leading order (mean field) approximation defined such that all physical quantities are real in this limit. 
It is demonstrated that the dynamics beyond mean field can be treated perturbatively. 
A detailed derivation of these perturbative corrections is presented in the appendix. 
The perturbative method is applied in Sec.~\ref{sec:numerics} to the PNJL model to investigate the effects of the complex phase of the action on different quantities. 
Here the expectation values of the Polyakov loop and its conjugate are of special interest.
In the present analysis, the split of the expectation values of the Polyakov loop and its conjugate at non-zero chemical potential arises once fluctuations of the fields are taken into account. 
Quantities like susceptibilities, in which mean field contributions partially or completely cancel, are sensitive to the corrections beyond mean field, as we shall discuss.
Finally in Sec.~\ref{sec:pions} further corrections beyond Hartree approximation are estimated, generated by propagating mesonic (quark-antiquark) modes. 
The lightest meson mode, the pseudoscalar pion with its approximate Nambu-Goldstone character, is the leading correction in this sector. 
Sec.~\ref{sec:conclusio} presents our conclusions and an outlook.

\section{The PNJL model}
\label{sec:pnjlmodel}

The two-flavour PNJL model including diquark degrees of freedom \cite{Roessner:2006xn} is derived from the Euclidean action
\begin{equation} 
{\cal S}_E(\psi, \psi^\dagger, \phi)= \int _0^{\beta=1/T} \intd\tau\int \intd^3x \left[\psi^\dagger\,\partial_\tau\,\psi + {\cal H}(\psi, \psi^\dagger, \phi)\right] + \delta{\cal S}_E(\phi,T)
\label{eqn:action}
\end{equation} 
with the fermionic Hamiltonian density \footnote{$\vec{\alpha} = \gamma_0\,\vec{\gamma}$ and $\gamma_4 = i\gamma_0$ in terms of the standard Dirac $\gamma$ matrices.}:
\begin{equation}
{\cal H} = -\iu\psi^\dagger\,(\vec{\alpha}\cdot \vec{\nabla}+\gamma_4\,m_0 -A_4)\,\psi + {\cal V}(\psi, \psi^\dagger)~,
\label{eqn:hamiltonian}
\end{equation}
where $\psi$ is the $N_\mathrm{f}=2$ doublet quark field and $m_0 = \mathrm{diag}(m_u,m_d)$ is the quark mass matrix. The quarks move in a background colour gauge field $A_4 = \iu A_0$, where $A_0 = \delta_{\mu 0}\,g{\cal A}^\mu_a\,t^a$ with the $\mathrm{SU}(3)_\mathrm{c}$ gauge fields ${\cal A}^\mu_a$ and the generators $t^a = \lambda^a/2$. The matrix valued, constant field $A_4$ relates to the (traced) Polyakov loop as follows:
\begin{equation}
\Phi=\frac{1}{N_\mathrm{c}}\tr_\mathrm{c}\,L \qquad\text{with }L= \exp\left(i\int_{0}^{\beta}
\intd\tau A_4\right)\quad\text{and }\beta = \frac{1}{T}~. \label{eqn:polyakovloop}
\end{equation}
In a convenient gauge (the so-called Polyakov gauge), the matrix $L$ is given a diagonal representation
\begin{equation}
\label{eqn:loopparametrization}
L = \exp \left[ \iu\,\left(\phi_3\,\lambda_3 +  \phi_8\,\lambda_8 \right) \right]~.
\end{equation}
The dimensionless effective fields $\phi_3$ and $\phi_8$ introduced here are identified with the Euclidean gauge fields in temporal direction, divided by temperature: $\phi_3={A_4^{(3)} }/{T}$ and $\phi_8={A_4^{(8)} }/{T}$.
These two fields are a parametrisation of the diagonal elements of $\mathrm{SU}(3)_{\mathrm{c}}$.
As such the ``angles'' $\phi_3$ and $\phi_8$ necessarily have to be real quantities in order to sustain the unitarity of the group.
An alternative parametrisation of the diagonal elements of $\mathrm{SU}(3)_{\mathrm{c}}$ is given by the Polyakov loop, $\Phi = \frac13 \tr_\mathrm{c}\,L$, and its conjugate, $\Phi^*= \frac13 \tr_\mathrm{c}\,L^\dagger$. 

The piece $\delta \mathcal{S}_\mathrm{E} = \frac{V}{T}\,\mathcal{ U}$ of the action (\ref{eqn:action}) carries information about the gluon dynamics. 
The potential $\mathcal{ U}$ effectively models the confinement-deconfinement transition and the region up to temperatures of roughly $T \lesssim 2\,T_{\mathrm{c}}$ in quarkless, pure gauge QCD on the mean field Ginzburg-Landau level. 
At temperatures very far above the transition a description of the thermodynamics with just the two order parameters $\Phi$ and $\Phi^*$ is not appropriate as transverse gluons will become important. 
Transverse gluon degrees of freedom cannot be described by Polyakov loops.

The Polyakov loop is an order parameter for confinement in $\mathrm{SU}(3)$ gauge theory.
In the confined low temperature phase the expectation value of the Polyakov loop vanishes, $\braket{\Phi} = 0$, while $\braket{\Phi} \neq 0$ implies deconfinement. 
Let $T_0$ be the critical temperature separating the two phases. 
As previously mentioned the symmetry which is restored at $T<T_0$ and broken above $T_0$ is the $\mathrm{Z}(3)$ centre-symmetry of $\mathrm{SU}(3)$\footnote{The centre of $\mathrm{SU}(3)$ contains all those $\mathrm{SU}(3)$ elements that commute with all other $\mathrm{SU}(3)$ elements, i.\,e.\ the elements $ \ee^{\iu\frac{2\pi}{3}k}\,\Eins$, with $k\in\mathds{Z}$ constituting a $\mathrm{Z}(3)$ subgroup of $\mathrm{SU}(3)$.}. 

Therefore, the Landau effective potential describing the dynamics, the Polyakov loop potential $\mathcal{U}(\Phi, T)$, has to be $\mathrm{Z}(3)$-symmetric in $\Phi$. 
The basic building blocks for such a potential are $\Phi^*\Phi$, $\Phi^3$ and ${\Phi^*}^3$ terms.
The potential used here differs from the simplest ansatz generating a first order phase transition as it is implemented in \cite{Ratti:2005jh}. Instead we use the ansatz given in \cite{Ratti:2006wg,Roessner:2006xn} motivated by the $\mathrm{SU}(3)$ Haar measure:
\begin{equation}
\frac{\mathcal{U}(\Phi,\,\Phi^*,\,T )}{T^4}=-\frac{1}{2}a(T)\,\Phi^*\Phi
+ b(T)\,\ln\left[1-6\,\Phi^*\Phi+4\left({\Phi^*}^3+\Phi^3\right)
-3\left(\Phi^*\Phi\right)^2\right]~,
\label{eqn:looppot}
\end{equation}
where the temperature dependent prefactors are given by
\begin{align}
a(T) & = a_0+a_1\left(\frac{T_0}{T}\right)
+a_2\left(\frac{T_0}{T}\right)^2 &\text{and}&& b(T) &=b_3\left(\frac{T_0}{T}
\right)^3.
\label{eqn:loopparam}
\end{align}
The logarithmic divergence near $\Phi^*,\,\Phi\to 1$ properly constrains the Polyakov loop to values attainable by the normalised trace of an element of $\mathrm{SU}(3)$.
The parameters of $\mathcal{U}(\Phi,\,\Phi^*,\,T )$ are chosen such that the critical temperature of the first order transition is indeed equal to $T_0$ (fixed at $270\mev$ \cite{Karsch:2000kv}) and that $\Phi^*\,,\Phi\to 1$ as $T\to\infty$. 

The numerical values using these constraints are taken as given in Refs.~\cite{Ratti:2006wg,Roessner:2006xn}
\begin{align}
a_0 &= 3.51\;, &a_1 &= -2.47\;, &a_2 &= 15.2\;, &b_3 &= -1.75\;. \nonumber
\end{align}
The resulting uncertainties are estimated to be about $6\,\%$ for $a_1$, less than $3\,\%$ for $a_2$ and  $2\,\%$ for $b_3$. 
The value $a_0 = \frac{16\pi^2}{45}$ chosen here reproduces the Stefan-Boltzmann limit. 
This is not mandatory, of course, since the high-temperature limit is governed by (transverse) gluonic degrees of freedom not covered by the Polyakov loop which represents the longitudinal gauge field. 
Alternative parametrisations of $\mathcal{U}$ are possible, such as the two-parameter form guided by the strong-coupling approach \cite{Fukushima:2003fw_Hatta:2003ga}, which has a different high temperature limit. 
In the present context these differences are not crucial as we systematically restrict ourselves to temperatures close to the transition region, $T\lesssim 2T_\mathrm{c}$, where different forms of $\mathcal{U}$ give remarkably similar results as pointed out in Ref.~\cite{Fukushima:2008wg}. 

\begin{figure}
\centering
\includegraphics[width=.65\textwidth]{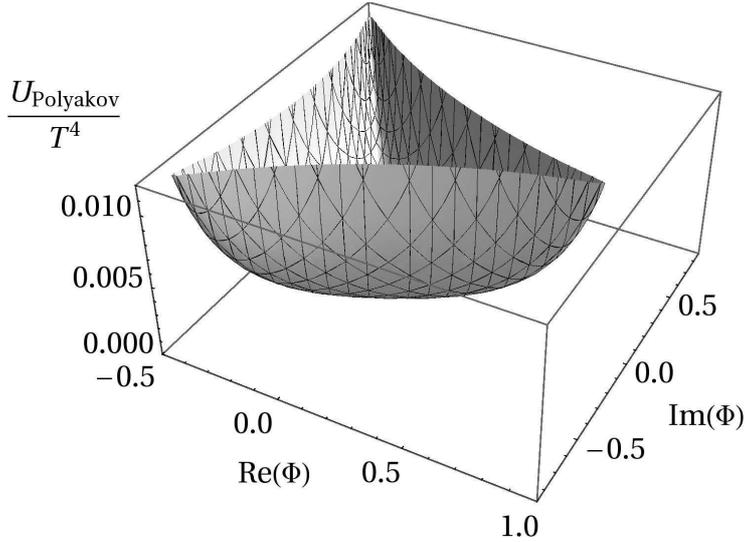}
\caption{The Polyakov loop potential ${\mathcal{U}(\Phi,\,\Phi^*,\,T )}/{T^4}$ plotted in the complex plane of $\Phi$ at $T = T_0 = 0.27\gev$. \label{fig:looppotential}}
\end{figure}

In Fig.~\ref{fig:looppotential} we plot the Polyakov loop potential using the parametrisation given in Refs.~\cite{Ratti:2006wg,Roessner:2006xn} at $T = T_0 = 0.27\gev$. 
This illustrates the $\mathrm{Z}(3)$ symmetry. 
The single minimum at $T<T_0$ becomes degenerate with three minima at $T=T_0$. 
Above $T_0$ only these three minima survive. 
Of course upon spontaneous breakdown of the $\mathrm{Z}(3)$ centre-symmetry, the three minima and the $\mathrm{Z}(3)$ centre-symmetry of the potential remain intact even though the vacuum expectation value does no longer show the symmetry of the potential. 

\smallskip
The NJL interaction term $\mathcal{V}$ in Eq.~(\ref{eqn:hamiltonian}) includes chiral $\mathrm{SU}(2)\times \mathrm{SU}(2)$ invariant four-point couplings of the quarks acting in pseudoscalar-isovector/scalar-isoscalar quark-antiquark and scalar diquark channels:
\begin{align}
\mathcal{V}= -\frac
{G}{2}\left[\left(\bar{\psi}\psi\right)^2+\left(\bar{\psi}\,i\gamma_5
\vec{\tau}\,\psi
\right)^2\right]
- \frac{H}{2}\left[\left(\bar{\psi}\,{\cal C}\gamma_5\tau_2\lambda_2
\,\bar{\psi}^{T}\right)\left(\psi^{T}\gamma_5\tau_2\lambda_2 {\cal C}
\,\psi\right)\right]~,
\label{eqn:potential}
\end{align}
where $\mathcal{C}$ is the charge conjugation operator.
These interaction terms in Eq.~(\ref{eqn:potential}) are obtained from a local colour current-current interaction between quarks, 
\begin{align}
\mathcal{L}_{\mathrm{int}} = - G_\mathrm{c}(\bar{\psi}\gamma_\mu t^a\psi)(\bar{\psi}\gamma^\mu t^a\psi)\;,
\nonumber
\end{align}
by a Fierz transformation which relates the coupling strengths $G$ and $H$ as $ G = \frac{4}{3}H$ which we choose not to alter\footnote{Additional terms generated by the Fierz transformation are of no importance in the present context and will be omitted.}.

The NJL model with two quark flavours is usually modelled with three parameters, a current quark mass $m_{u,d}$, a local four quark coupling strength $G$ and a three-momentum cutoff $\Lambda$. The parameters used here are the ones used in \cite{Ratti:2005jh,Ratti:2006wg,Roessner:2006xn}:
\begin{align}
m_{u,d} &= 5.5\mev\;,&G &= \frac{4}{3}H = 10.1\gev^{-2}\;,& \Lambda &= 0.65\gev\;,
\nonumber
\end{align}
fixed to reproduce the pion mass and decay constant in vacuum and the chiral condensate as $m_\pi =$ 139.3 MeV, $f_\pi =$ 92.3 MeV and $\langle\bar{\psi}_u\psi_u\rangle = - (251$ MeV)$^3$.

To evaluate the thermodynamic properties of the model the quark degrees of freedom are integrated out.
New auxiliary fields are introduced by bosonisation, absorbing quark-antiquark and quark-quark (antiquark-antiquark) correlations. 
These are a scalar-pseudoscalar field $(\sigma,\,\vec{\pi}\,)$ and a diquark (antidiquark) field $\Delta$ ($\Delta^*$). 
The resulting thermodynamic potential then reads
\begin{equation}
\Omega_0 = \frac{T}{V} \, \mathcal{S}_{\mathrm{bos}}= \mathcal{U}\left(\Phi,\,\Phi^*,\,T\right)-\frac{T}{2}\sum_n\int\frac{\intd^3p}
{\left(2\pi\right)^3}\Tr \ln\left[\beta \tilde{S}^{-1}\left(i\omega_n,\vec{p}\,\right)
\right]+\frac{\sigma^2}{2G}+\frac{\Delta^*\Delta}{2H}~,
\label{eqn:omegageneral}
\end{equation}
where the Matsubara sum runs over $\omega_n=(2n+1)\pi\,T$ reproducing antiperiodic boundary conditions in the Euclidean time direction.
The inverse Nambu-Gor'kov propagator $\tilde{S}^{-1}$ in Eq.~(\ref{eqn:omegageneral}) is defined by
\begin{equation}
\tilde{S}^{-1}\left(i\omega_n,\vec{p}\,\right)=\left({{\begin{array}{ccc} 
i\gamma_0\,\omega_n-\vec{\gamma}\cdot\vec{p}-
m+\gamma_0\left(\mu - \iu A_4 \right)& \Delta\gamma_5\tau_2\lambda_2\\
-\Delta^*\gamma_5\tau_2\lambda_2& 
i\gamma_0\, \omega_n-\vec{\gamma}\cdot\vec{p}-
m-\gamma_0\left(\mu - \iu A_4 \right)
\end{array}}}\right).
\label{eqn:ngprop}
\end{equation}
The mass of the quark-quasiparticles is given as in the standard NJL model by the gap equation
\begin{equation}
m=m_0-\langle\sigma\rangle=m_0-G\langle\bar{\psi}\psi\rangle\;.
\label{eqn:mass}
\end{equation}

The Matsubara sum is evaluated analytically. The quasiparticle energies emerging in this procedure are related to the solutions of $\det\big[\tilde{S}^{-1}(p_0)\big] = 0$. %, which can be mapped onto an eigenvalue problem. 
The bosonised action then reads
\begin{multline}
\Omega_0 = \frac{T}{V} \, \mathcal{S}_{\mathrm{bos}} =  \mathcal{U}\left(\Phi,\,\Phi^*,\,T\right)+\frac{\sigma^2}{2G}+
\frac{\Delta^*\Delta}{2H}\\
 - 2N_f\int\frac{\intd^3p}{\left(2\pi\right)^3}\sum_j \left\{
T\ln\left[1+\ee^{-E_j/T}\right]+\frac12 \Delta E_j
\right\}\;,
\label{eqn:bosaction}
\end{multline}
with six distinct quasiparticle energies
\begin{align}
E_{1,2}&=\varepsilon(\vec{p}\,) \pm \tilde{\mu}_b~,
\nonumber\\
E_{3,4}&=\sqrt{(\varepsilon(\vec{p}\,)+\tilde{\mu}_r)^2+|\Delta|^2}\pm i\,T\,\phi_3~,
\nonumber\\
E_{5,6}&=\sqrt{(\varepsilon(\vec{p}\,)-\tilde{\mu}_r)^2+|\Delta|^2}\pm i\,T\,\phi_3~,
\end{align}
where $\varepsilon(\vec{p}\,) = \sqrt{\vec{p\,}^2+m^2}$. Additionally we have introduced
\begin{align}
\tilde{\mu}_b&=\mu+2i\,T\,\frac{\phi_8}{\sqrt{3}}\;, & \tilde{\mu}_r&=\mu-i\,T\,\frac{\phi_8}{\sqrt{3}}\;.
\end{align}
The energy difference $\Delta E_j$ is defined as the difference of the quasiparticle energy and the energy of a free fermion, $\varepsilon_0 = \sqrt{\vec{p\,}^2+m_0^2}$: $\Delta E_j = E_j - \varepsilon_0 \pm \mu$. 
The form of the bosonised action, Eq.~(\ref{eqn:bosaction}), does not allow to factor out the Polyakov loop fields $\Phi$ and $\Phi^*$, as it was done in Ref.~\cite{Ratti:2005jh}.
Instead we keep the form of Eq.~(\ref{eqn:bosaction}) using $\phi_3$ and $\phi_8$ with $\phi_3,\,\phi_8\in \mathds{R}$.\footnote{As the parameter space of $\phi_3$ and $\phi_8$ is periodic there are different parameter sets representing the same physics. We use the (triangular shaped) domain $\lbrace ( \phi_8 \geqq -\frac{\pi}{\sqrt{3}} ) \wedge (\phi_8 \leqq \sqrt{3}(\phi_3+\frac{2\pi}{3}) ) \wedge (\phi_8 \leqq \sqrt{3}(-\phi_3+\frac{2\pi}{3}) ) \rbrace $. Note that the periodic domain of $L$ and $L^\dagger$ is $3!$-times larger than the domains for $\Phi$ and $\Phi^*$ (or equivalently $\phi_3$ and $\phi_8$) due to the trace's invariance under unitary transformations of $L$.}

\medskip
The introduction of the Polyakov loop outlined above formally leads to a complex valued action as soon as $\mu\neq0$. 
This phenomenon is usually called fermion sign problem. 
Due to the connection of $\phi_3$ and $\phi_8$ to the QCD colour gauge group $\mathrm{SU}(3)_{\mathrm{c}}$, we must  require $\phi_3$ and $\phi_8$ to be real fields at all times. 
The mean field approximation, $\Omega_{\mathrm{MF}}$, of the thermodynamic potential $\Omega$ must be introduced such that it satisfies this constraint imposed by the gauge group. 
Identification up to a constant of $\Omega_{\mathrm{MF}}$ with the (real) pressure $p$ in this approximation then requires that non-Hermitian structures of the inverse quasiparticle quark propagator do not contribute. 
One way to establish such a lowest order approximation is to use the real part of the thermodynamic potential in the mean field equations. 

The necessary condition for the minimisation of the effective action in a standard situation is, in general, 
\begin{equation}
\frac{\partial\,\Omega}{\partial\theta_i}=0 \;, 
\label{eqn:saddle}
\end{equation}
where $\theta_i$ stands for the fields representing the relevant degrees of freedom (in our case: $\theta = \left( \sigma,\,\Delta,\,\phi_3 ,\,\phi_8 \right)$). 
In order to always comply with $\phi_3,\,\phi_8\in \mathds{R}$ 
we define the mean field thermodynamic potential, with $\Omega_0$ of Eq.~(\ref{eqn:bosaction}), by
\begin{equation}
\label{eqn:omegamf}
\Omega_{\mathrm{MF}} =\Rens\left[ \,\Omega_0\right] = \Rens\left[\frac{T}{V}\,\mathcal{S}_{\mathrm{bos}}\right]~.
\end{equation}
The mean field equations then read
\begin{equation}
\frac{\partial\,\Omega_{\mathrm{MF}}}{\partial \left( \sigma, \Delta, \phi_3, \phi_8\right)}=\frac{\partial\,\Rens\left[ \,\Omega_{0}\right]}{\partial \left( \sigma, \Delta, \phi_3, \phi_8\right)}=0.
\label{eqn:mfeqn}
\end{equation}
The hereby neglected imaginary part of this derivative, $\frac{\partial\,\Im[\,\Omega_{0}]}{\partial \left( \sigma, \Delta, \phi_3, \phi_8\right)}$, will be taken into account by writing down a series in powers of this residual gradient. 
In addition it is also possible to correct for deviations of the potential from a gaussian shape which is assumed for the mean field approximation. 
As explained in the appendix it is most convenient to consider both types of corrections simultaneously using Feynman graphs to construct all possible terms. 

\medskip

\begin{figure}
\centering
\includegraphics[width=.65\textwidth]{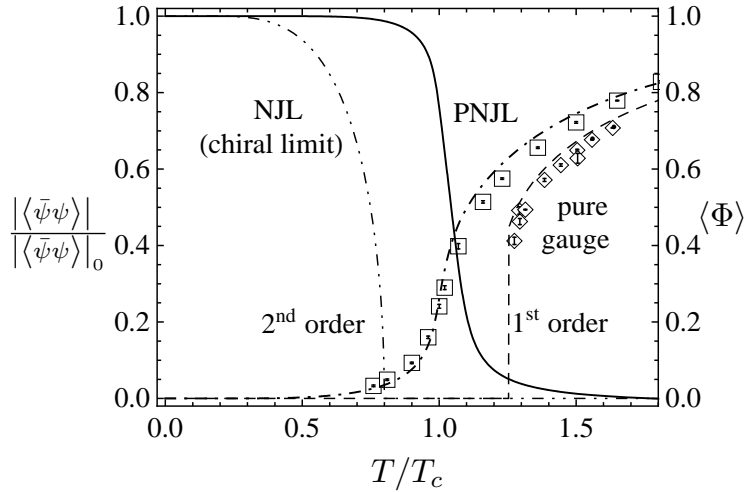}
\caption{Chiral condensate normalised to its value at temperature $T=0$ (dash-double-dotted line) in the NJL model with massless quarks, and Polyakov loop $\braket{\Phi}$ in the pure gauge model (dashed line). 
The PNJL model (with non-zero quark masses) shows dynamical entanglement of the chiral (solid line) and Polyakov loop (dash-dotted line) crossover transitions. 
For comparison lattice data for the Polyakov loop  in pure gauge and full QCD (including quarks) are also shown \cite{Kaczmarek:2005ui} \label{fig:loopwithquarks}.} 
\end{figure}

A variety of PNJL model results (equations of state, phase diagrams, susceptibilities) have been obtained in previous  calculations \cite{Ratti:2006wg,Roessner:2006xn,Ratti:2007jf} based on the mean field equations (\ref{eqn:mfeqn}). 
At this point it is instructive to examine how chiral and Polyakov loop dynamics cooperate to produce crossover transitions (at zero chemical potential) which end up in a narrow overlapping range of temperatures (see Fig.~\ref{fig:loopwithquarks}). 
In isolation, the pure gauge Polyakov loop sector and the NJL sector in the chiral limit show first (second) order phase transitions with critical temperatures far separated, as demonstrated by the dashed (dash-double dotted) lines in Fig.~\ref{fig:loopwithquarks}. 
When entangled in the PNJL model, these transitions (with non-zero quark masses) move together to form a joint crossover pattern. 

\section{Fluctuations and corrections beyond mean field}
\label{sec:corr}

The discussion of fluctuation corrections to the mean field approximation has two parts:
\begin{itemize}
 \item corrections arising from the imaginary part of $\Omega_0$ and involving the Polyakov loop $\Phi$ and its complex conjugate $\Phi^*$;
 \item corrections from dynamical fluctuations involving propagating meson fields, with emphasis on the pion. 
\end{itemize}
The first item is the primary topic of this present section. 
The second item will be relegated to a separate Section~\ref{sec:pions}. 

The mean field equations (\ref{eqn:mfeqn}) establish a leading order approximation satisfying the reality constraints on the fields $\phi_3$ and $\phi_8$. 
Eq.~(\ref{eqn:mfeqn}) is at the same time the necessary condition for the maximisation of the modulus $\left\vert \ee^{-\mathcal{S}_{\mathrm{E}}} \right\vert$ of the thermodynamic weight in the path integral\footnote{In analogy to the procedure in Minkowskian space-time one might argue that the complex phase needs to become stationary. Taking the thermodynamic limit one observes that the stationary phase field configuration is favoured over any other configuration by the factor $\frac{V}{T}$, while the absolute value is favoured over other configurations by a factor $\ee^{ \frac{V}{T}}$.}:

The direct connection $\Phi = \Phi(\phi_3,\,\phi_8)$ of the two parametrisations, ($\phi_3$, $\phi_8$) on the one hand and ($\Phi$, $\Phi^*$) on the other hand, is lost once we step away from mean field and calculate thermodynamic expectation values: $\braket{\Phi}\neq\Phi(\braket{\phi_3},\,\braket{\phi_8})$.\footnote{
Recall that the thermal expectation value $\braket{\cdots}$ is a weighted sum (integral) over various thermal field configurations. 
The equality $\braket{\Phi} = \Phi(\braket{\phi_3},\,\braket{\phi_8})$ holds only if $\Phi$ and $\Phi^*$ are linear functions of $\phi_3$ and $\phi_8$. 
This is not the case. }
This observation is crucial when comparing the present method of approximation to schemes in previous publications \cite{Ratti:2005jh,Sasaki:2006ww,Ghosh:2006qh,Mukherjee:2006hq,Zhang:2006gu}. 
In these publications the fields $\phi_3$, $\phi_8$ have been replaced by $\Phi$, $\Phi^*$ before doing mean field approximation. 
This implies that $\braket{\Phi}$ and $\braket{\Phi^*}$ (and \emph{not} $\Phi$, $\Phi^*$) are treated as independent mean field degrees of freedom. 
The minimisation of $\Omega_0$ is then performed requiring that $\braket{\Phi}$ and $\braket{\Phi^*}$ are real quantities. 
In such approximation schemes it is therefore not possible to find a way back to the (real) quantities $\phi_3$, $\phi_8$: $\braket{\Phi}$ and $\braket{\Phi^*}$ already comprise fluctuations of $\phi_3,\, \phi_8\in\mathds{R}$. 
In other words, the definition of the lowest order approximation (which is usually referred to as mean field approximation) is different in Refs.~\cite{Ratti:2005jh,Sasaki:2006ww,Ghosh:2006qh,Mukherjee:2006hq,Zhang:2006gu} and this work. 
The definition of the lowest order (mean field) approximation in this work allows to strictly separate contributions originating in constant and fluctuating parts of the fields. 

The combination of the constraints $\Omega\in\mathds{R}$ and $\phi_3,\,\phi_8\in\mathds{R}$ allow only certain limited configurations of $\phi_3$ and $\phi_8$.
In mean field approximation the condition $\phi_3,\,\phi_8\in\mathds{R}$ implies that, due to Eq.~(\ref{eqn:loopparametrization}), $\Phi$ and $\Phi^*$ are the complex conjugates and we find $\braket{\Phi}_{\mathrm{MF}} = \Phi_{\mathrm{MF}}$, $\braket{\Phi^*}_{\mathrm{MF}} = \Phi^*_{\mathrm{MF}}$. 
At $\mu=0$ the Polyakov loop $\Phi$ and its complex conjugate $\Phi^*$ are treated equally due to charge conjugation invariance. 
It follows that $\Phi_{\mathrm{MF}} = \Phi^*_{\mathrm{MF}} \in \mathds{R}$ in mean field approximation, fixing $\phi_8=0$.
The Polyakov loop effective potential $\mathcal{U} = \mathcal{U}(T,\,\Phi,\,\Phi^*)$ in its parametrisation (\ref{eqn:looppot}) is minimal for $\Phi = \Phi^*$ at fixed $\vert \Phi \vert$.
We find that the Polyakov loop potential $\mathcal{U}$ is always strong enough to keep $\Phi = \Phi^*$ or, equivalently, $\phi_8=0$.
Not all parametrisations of $\mathcal{U}$ will maintain this solution. 
If the curvature of the potential $\mathcal{U}$ is not strong enough, the solution $\phi_8=0$ becomes instable, and $\phi_8=0$ is the position of a local maximum of the potential.
In Ref.~\cite{Sasaki:2006ww} a symptom of this fact has been described: the susceptibility associated with $\Re \Phi$ may become negative. 
The potential used in the present work does not show such deficiencies. 

After these preparatory remarks we proceed to develop a calculational scheme which systematically treats corrections to the mean field approximation, Eqs.~(\ref{eqn:omegamf}) and (\ref{eqn:mfeqn}). 
The basic idea is, as usual, to expand the thermodynamic potential $\Omega$ around its mean field limit $\Omega_{\mathrm{MF}}$. 
Technical details of the derivation are summarized in the appendix. 
The result including next-to-leading order is
\begin{equation}
\label{eqn:numOmega}
\Omega %= \Omega_{0} \;-\;\; \frac12\,\frac{T}{V}\;\;\parbox{25pt}{\includegraphics{./graph9.eps}} 
= \Omega_{\mathrm{MF}} - \left. \frac12 \left( \frac{\partial \Omega_{0}}{\partial \theta}  \right)^T  \cdot \left[ \frac{\partial^2 \Omega_{0}}{\partial \theta^2} \right]^{-1} \cdot  \frac{\partial \Omega_{0}}{\partial \theta}  \right\vert_{\theta= \theta_{\mathrm{MF}}}~,
\end{equation}
starting from the (complex) $\Omega_0$ of Eq.~(\ref{eqn:bosaction}), with $\Omega_{\mathrm{MF}} $ defined by Eq.~(\ref{eqn:omegamf}), and with $\theta_{\mathrm{MF}}$ determined by Eq.~(\ref{eqn:mfeqn}). 
The gradients $\partial \Omega_0 / \partial \theta$ are understood with the set of field variables $\theta = (\theta_i) = ( \sigma, \Delta, \phi_3, \phi_8)$ arranged in vector form, and $\partial^2 \Omega_0 / \partial \theta^2$ stands for the matrix $(\partial^2 \Omega_0 / \partial \theta_i \partial \theta_j)$. 
The correction term in (\ref{eqn:numOmega}) is taken using the mean field configuration, $ \theta = \theta_{\mathrm{MF}}$. 
Note that this term takes care of the contributions from $\Im \Omega_0$ in such a way that $\Omega$ remains a real quantity. 

The thermal expectation value $\braket{f}$ of a physical quantity $f$ is calculated according to
\begin{equation}
\label{eqn:numfield}
\braket{f} %= f(\theta_{\mathrm{MF}}) \;+\;\;\parbox{25pt}{\includegraphics{./graph10.eps}}
 = f(\theta_{\mathrm{MF}}) - \left. \left( \frac{\partial \Omega_{0}}{\partial \theta}  \right)^T\cdot \left[ \frac{\partial^2 \Omega_{0}}{\partial \theta^2} \right]^{-1} \cdot \frac{\partial f}{\partial \theta} \right\vert_{\theta= \theta_{\mathrm{MF}}} \;.
\end{equation}
Applications will now be given, in particular, for the Polyakov loop and for susceptibilities.

\section{Results}
\label{sec:numerics}

The numerical calculations presented in this section are performed in the thermodynamic limit, i.\,e.\ in leading order of the $\frac{T}{V}$-expansion, and up to first order in the $\delta$-expansion, as explained in detail in the appendix. 
The thermodynamic potential is determined by Eq.~(\ref{eqn:numOmega}). 
Thermal expectation values are computed using Eq.~(\ref{eqn:numfield}).

\subsection{The Polyakov loop \boldmath{$\braket{\Phi}$} and its conjugate \boldmath{$\braket{\Phi^*}$}}

With the mean field definition (\ref{eqn:omegamf}) the Polyakov loop expectation values $\braket{\Phi}$ and $\braket{\Phi^*}$ turn out to be equal in this limit, given the reality constraint on $\Omega_{\mathrm{MF}}$. 
It is the corrections from $\Im \Omega_0$ induced by the temporal gauge fields which cause the splitting of $\braket{\Phi}$ and $\braket{\Phi^*}$. 

\begin{figure}
\begin{minipage}{.48\textwidth}
\includegraphics[width=\textwidth]{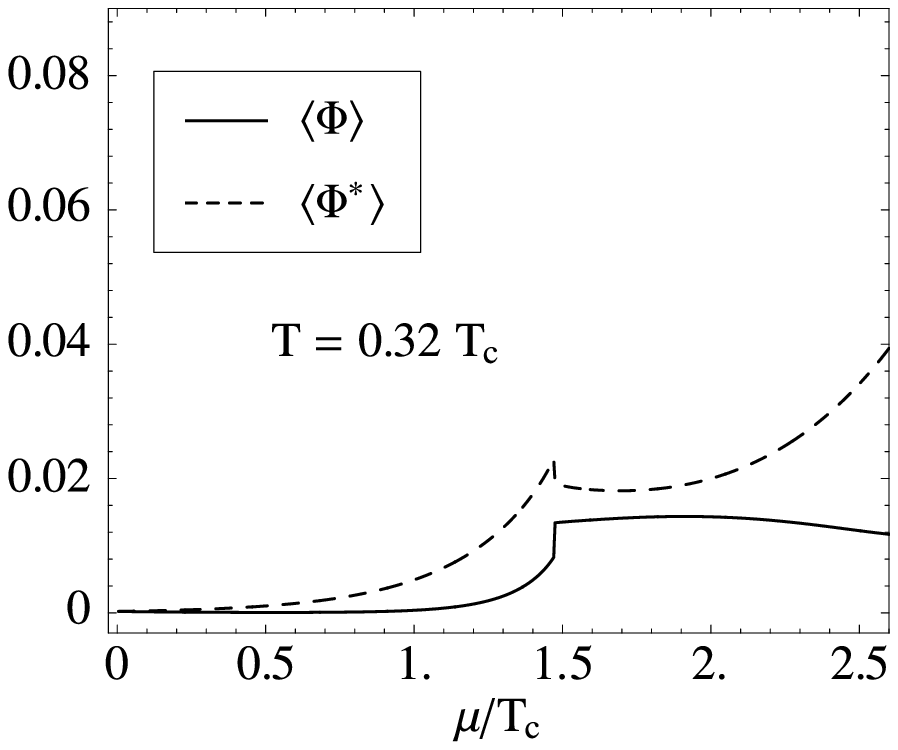}
\end{minipage}\hfill
\begin{minipage}{.48\textwidth}
\includegraphics[width=\textwidth]{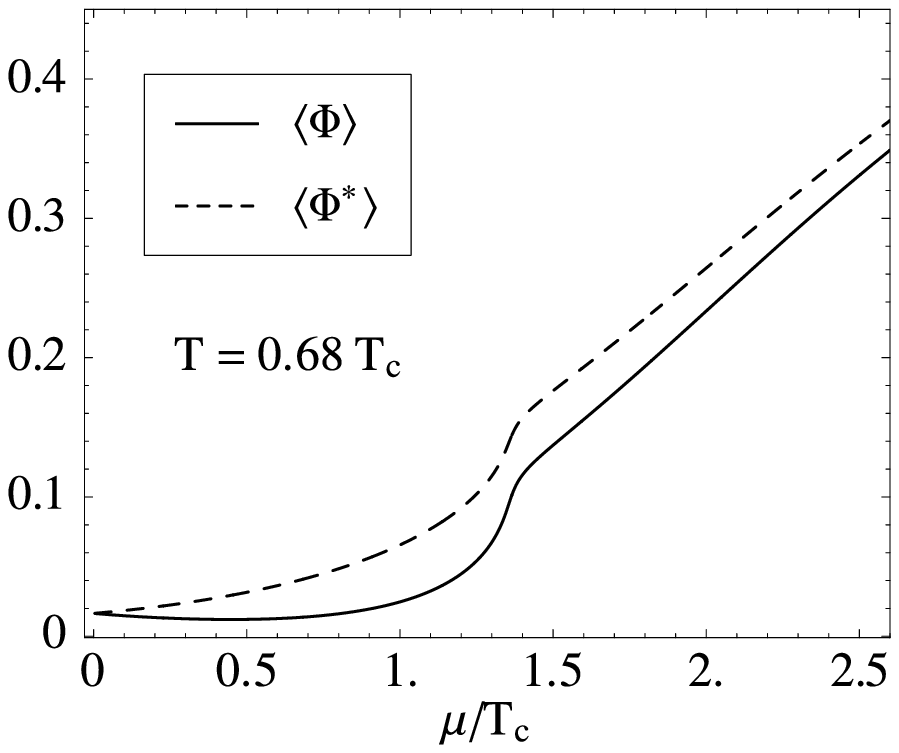}
\end{minipage}\\[6pt]
\begin{minipage}{.48\textwidth}
\includegraphics[width=\textwidth]{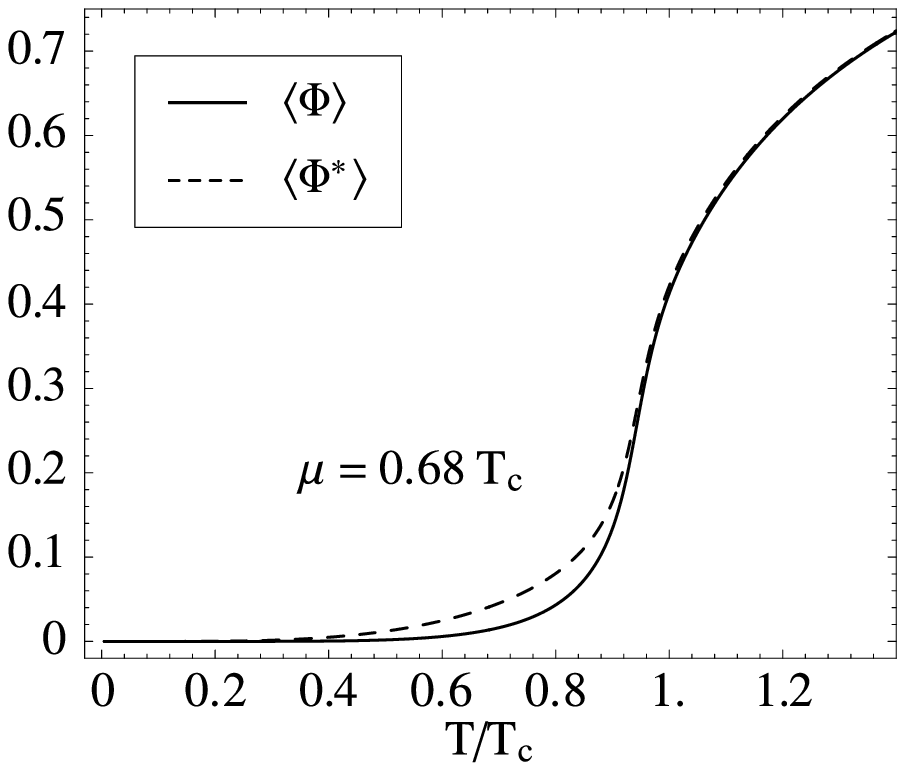}
\end{minipage}\hfill
\begin{minipage}{.48\textwidth}
\includegraphics[width=\textwidth]{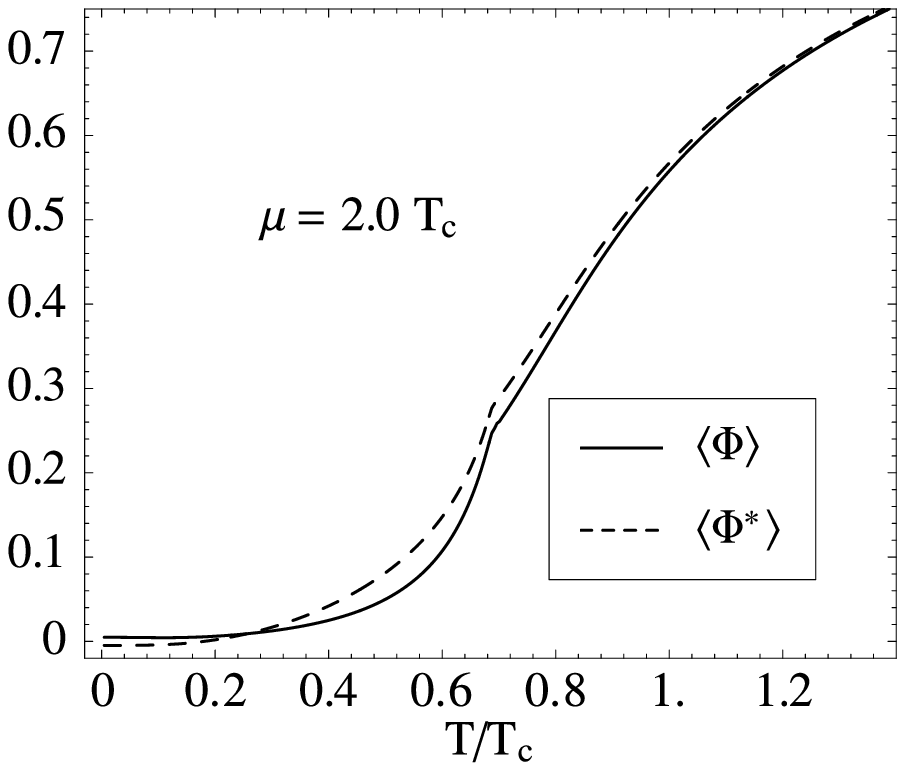}
\end{minipage}
\caption{\label{fig:loop}
Examples of thermal expectation values of the Polyakov loop $\braket{\Phi}$ and its conjugate $\braket{\Phi^*}$. In the upper row $\braket{\Phi}$ and $\braket{\Phi^*}$ are plotted as functions of the chemical potential $\mu$ at constant temperature $T$. Below $\braket{\Phi}$ and $\braket{\Phi^*}$ are plotted as functions of temperature $T$ at constant chemical potential $\mu$.}
\end{figure}

The difference $\braket{\Phi^*} - \braket{\Phi}$ vanishes at zero quark chemical potential $\mu$ and has the same sign as $\mu$, in agreement with results of Ref.~\cite{Dumitru:2005ng}. 
As can be seen from Fig.~\ref{fig:loop} the difference $\braket{\Phi^*} - \braket{\Phi}$ is pronounced around the phase transitions. 
In the upper left panel of Fig.~\ref{fig:loop} the influence of the first order phase transition separating the chiral and the diquark phase at low temperature can be seen as a jump in both $\braket{\Phi}$ and $\braket{\Phi^*}$. 
The second order phase transition separating the diquark regime from the high temperature quark-gluon phase can be identified as a kink in the lower right panel of Fig.~\ref{fig:loop}.

\subsection{Susceptibilities and phase diagram}

A susceptibility $\chi_g$ involving a quantity $g$ is defined by 
\begin{equation}
\label{eqn:genericsusc}
\chi_g^2 = V\,\braket{(g-\braket{g})^2} = V\left(\braket{g^2}-\braket{g}^2\right)~. 
\end{equation}
Susceptibilities of special interest in the present context are the ones related to the dynamical quark mass, $m=m_0-\sigma$, and to the Polyakov loop. 
They are expressed in terms of the inverse matrix of the second derivatives of the full thermodynamic potential $\Omega$:
\begin{align}
\label{eqn:chi_m}
 \chi_M^2 &= V\left(\braket{m^2}-\braket{m}^2\right)= \;T\,\left[\frac{\partial^2\Omega}{\partial \theta_i\partial \theta_j}\right]^{-1}_{m,\,m} \\
\label{eqn:chi_phi}
 \qquad\qquad\chi_{\Phi}^2 &= V\left(\braket{{\Phi}^2}-\braket{{\Phi}}^2\right) \;=\; T\,\left[\frac{\partial^2\Omega_{\mathrm{full}}}{\partial \theta_i\partial \theta_j}\right]^{-1}_{\Phi,\,\Phi}\qquad\qquad\\
\label{eqn:chi_rephi}
 \qquad\qquad\chi_{\Re \Phi}^2 &= \; \frac{T}{4}\,\left[\frac{\partial^2\Omega}{\partial \theta_i\partial \theta_j}\right]^{-1}_{\Phi,\,\Phi} + \frac{T}{2}\,\left[\frac{\partial^2\Omega_{\mathrm{full}}}{\partial \theta_i\partial \theta_j}\right]^{-1}_{\Phi,\,\Phi^*} + \frac{T}{4}\,\left[\frac{\partial^2\Omega}{\partial \theta_i\partial \theta_j}\right]^{-1}_{\Phi^*,\,\Phi^*} ~.
\end{align}
These susceptibilities are calculated using the graph rules given in Tab.~\ref{tab:chisqr} of the appendix which lead to the following explicit form: 
\begin{multline}
\label{eqn:gausssus}
\chi^2_g = T  \left. \left( \frac{\partial g}{\partial \theta}  \right)^T  \cdot \left[ \frac{\partial^2 \Omega_{0}}{\partial \theta^2} \right]^{-1} \cdot  \frac{\partial g}{\partial \theta}  \right\vert_{\theta= \theta_{\mathrm{MF}}} \\
-2\,T \left.\left( \frac{\partial g}{\partial \theta}  \right)^T  \cdot \left[ \frac{\partial^2 \Omega_{0}}{\partial \theta^2} \right]^{-1} \cdot 
\frac{\partial^2 g}{\partial \theta^2} \cdot 
\left[ \frac{\partial^2 \Omega_{0}}{\partial \theta^2} \right]^{-1}\cdot 
\frac{\partial \Omega_{0}}{\partial \theta} \right\vert_{\theta= \theta_{\mathrm{MF}}} \qquad\qquad\rule{0pt}{0pt} \\
+ T\,\sum_{i,j,k} \frac{\partial^3 \Omega_{0}}{\partial \theta_i\partial \theta_j\partial \theta_k} 
\left(\left[ \frac{\partial^2 \Omega_{0}}{\partial \theta^2} \right]^{-1}\cdot 
\frac{\partial g}{\partial \theta} \right)_i 
\left(\left[ \frac{\partial^2 \Omega_{0}}{\partial \theta^2} \right]^{-1}\cdot 
\frac{\partial g}{\partial \theta} \right)_j \\
\left. \times \left(\left[ \frac{\partial^2 \Omega_{0}}{\partial \theta^2} \right]^{-1}\cdot 
\frac{\partial \Omega_{0}}{\partial \theta} \right)_k \right\vert_{\theta= \theta_{\mathrm{MF}}} ~.%\quad+\;\; O(\beta = 2) + O(\alpha = 2)
\end{multline}
Here $g$ stands for $m$ or $\Phi$, respectively. 
The first term in Eq.~(\ref{eqn:gausssus}) is the susceptibility of the gaussian theory whereas the other terms are interpreted as corrections. 

The susceptibilities $\chi_M$ and $\chi_\Phi$ serve as indicators for boundaries between phases when drawing a phase diagram in the plane of temperature and chemical potential. 
For smooth crossover transitions, such boundaries are not rigorously defined. 
Several criteria can be used to determine a transition line separating the region of spontaneously broken chiral symmetry from the quark-gluon phase. 
We use here the maxima of the chiral susceptibilities $\chi_M$ and of the Polyakov loop susceptibility $\chi_{\Re\Phi}$ in comparison with the maximal slopes $\intd m/\intd T$ and $\intd\braket{\Phi^*+\Phi} / \intd T$ of the corresponding quantities which act, respectively, as order parameters in the limiting situations of exact chiral $ \mathrm{SU}(2) \times \mathrm{SU}(2) $ symmetry or $\mathrm{Z}(3)$ symmetry. 
 
\begin{figure}
\begin{minipage}{.48\textwidth}
\includegraphics[width=\textwidth]{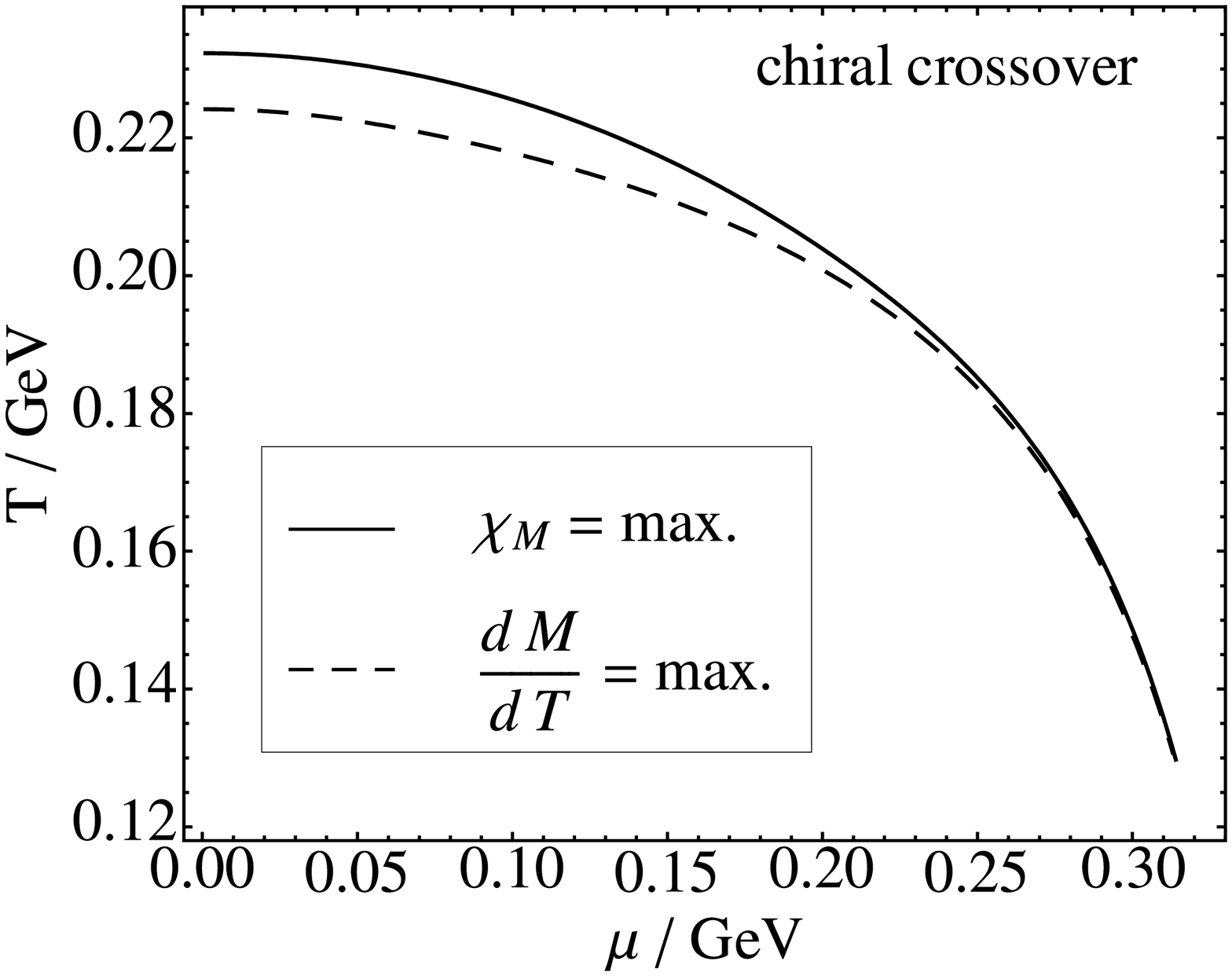}
\end{minipage}\hfill
\begin{minipage}{.48\textwidth}
\includegraphics[width=\textwidth]{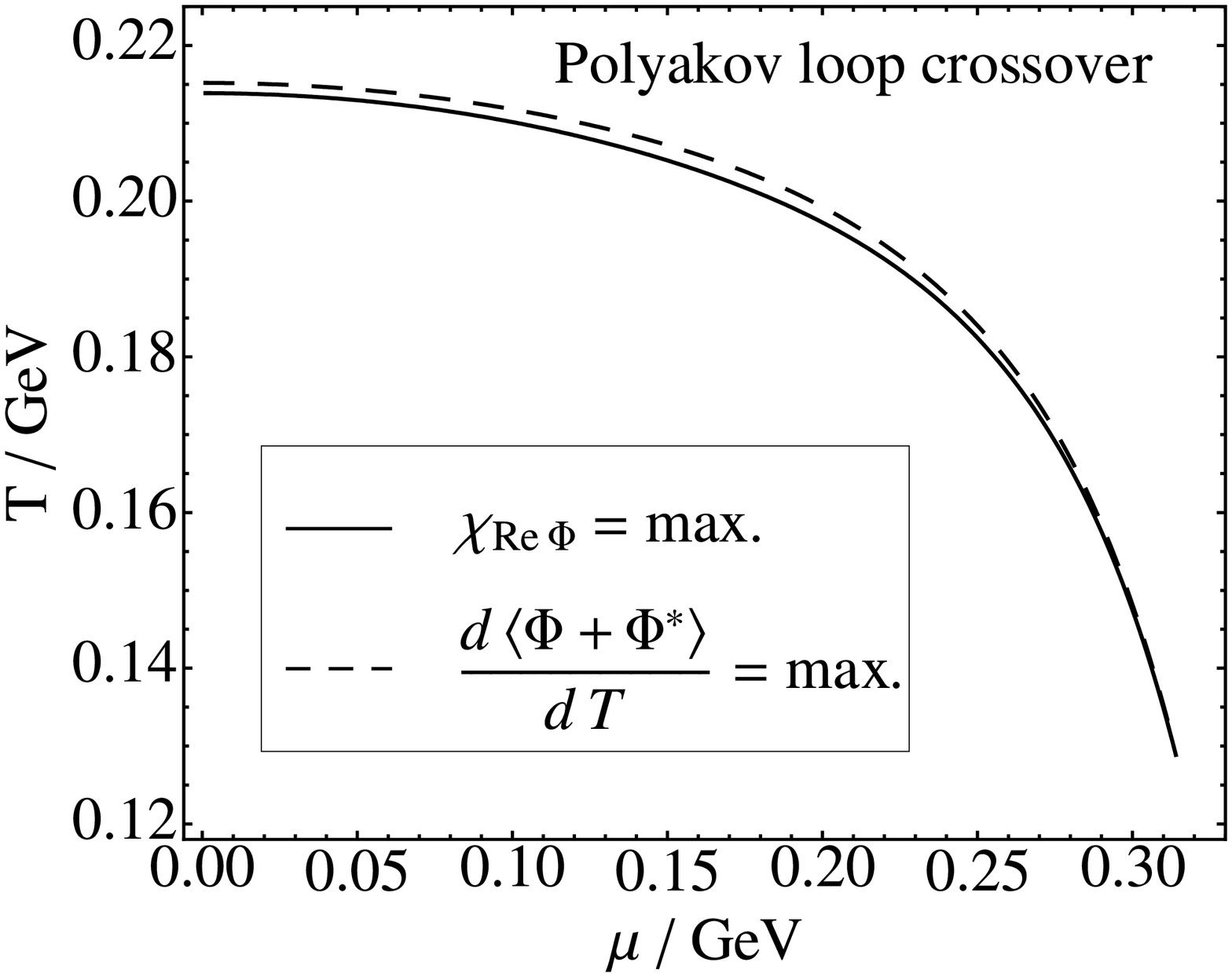}
\end{minipage}
\caption{
\label{fig:sustemp}
Comparison of the transition lines obtained by determination of the maximum in the chiral susceptibility (left panel solid line) and the Polyakov loop susceptibility $\chi_{\Re \Phi}$ (right panel solid line) with the transition lines fixed by the maximal change with respect to temperature of constituent quark mass (left panel dashed line) and average of the real part of the Polyakov loop $\frac12\braket{\Phi+\Phi^*}$ (right panel dashed line).}
\end{figure}

Fig.~\ref{fig:sustemp} shows a comparison of crossover transition lines found with the two criteria just mentioned.
As both criteria are linked via the quadratic term in the action, all curves finally converge to the same point, the critical point (here in the absence of diquark condensation).
A singularity in the second derivative of the action (or equivalently in the propagator) enforces this unique intersection point where the specific heat and other quantities show singular behaviour.

Comparing our Fig.~\ref{fig:sustemp} with corresponding results in other publications (see Fig.~16 in Ref.~\cite{Sasaki:2006ww} and Fig.~4 in Ref.~\cite{Abuki:2008nm}) one finds that the detailed behaviour of the deconfinement crossover transition line depends sensitively on the parameter choice and regularisation prescription. 
In the present case of a strong coupling a joint course of chiral and deconfinement crossover line is observed. 
When the coupling becomes weaker (e.\,g.\ due to parameter choice and regularisation prescription as in Refs.~\cite{Sasaki:2006ww,Abuki:2008nm}) the transition lines may deviate. 
In any case one should note that such deviations appear at large chemical potentials approaching the typical cutoff scale of the model. 
Any conclusions drawn at such energy or momentum scales should be handled with care. 

\begin{figure}
\begin{minipage}{.47\textwidth}
\includegraphics[width=\textwidth]{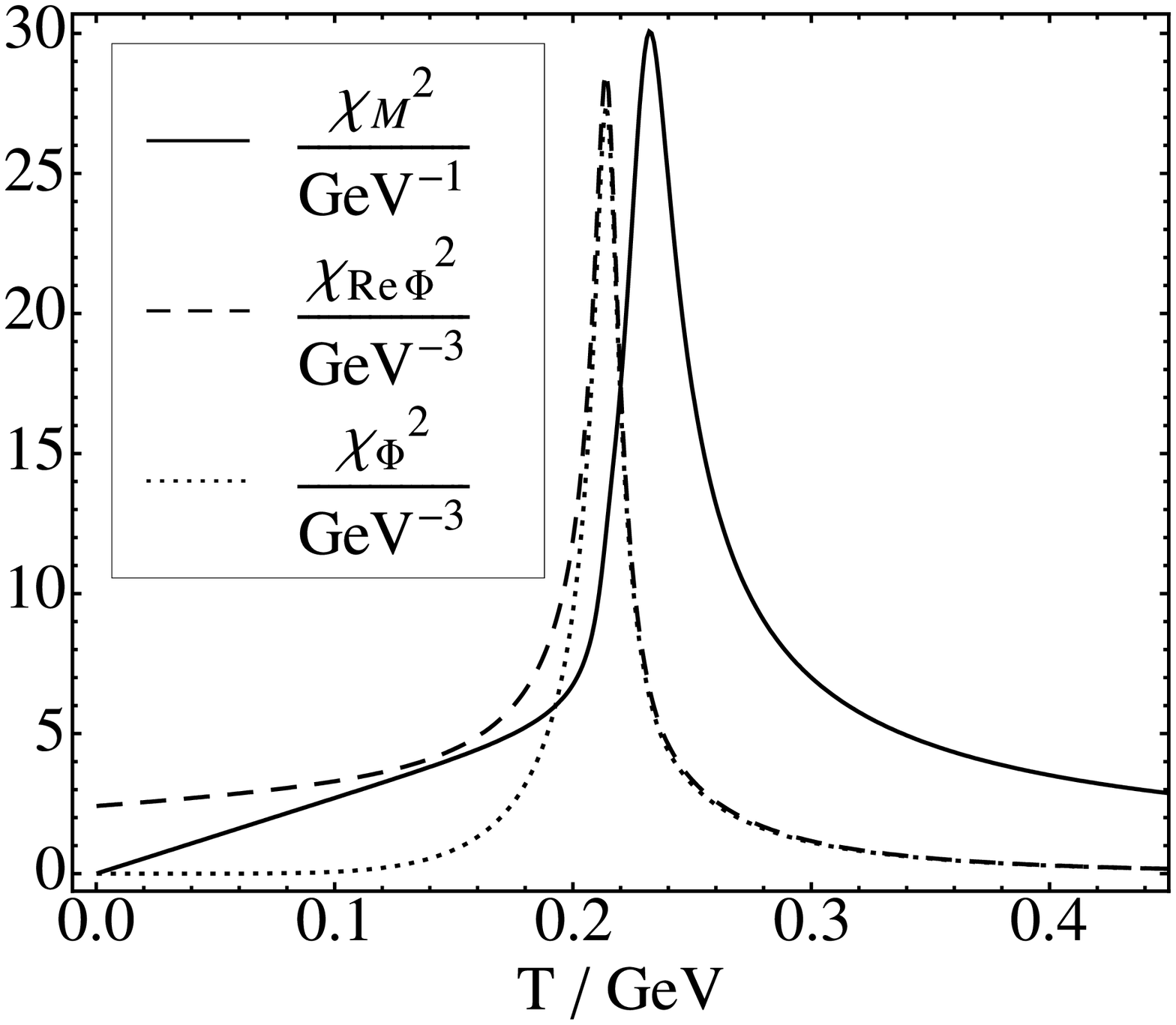}
\end{minipage}\hfill
\begin{minipage}{.47\textwidth}
\includegraphics[width=\textwidth]{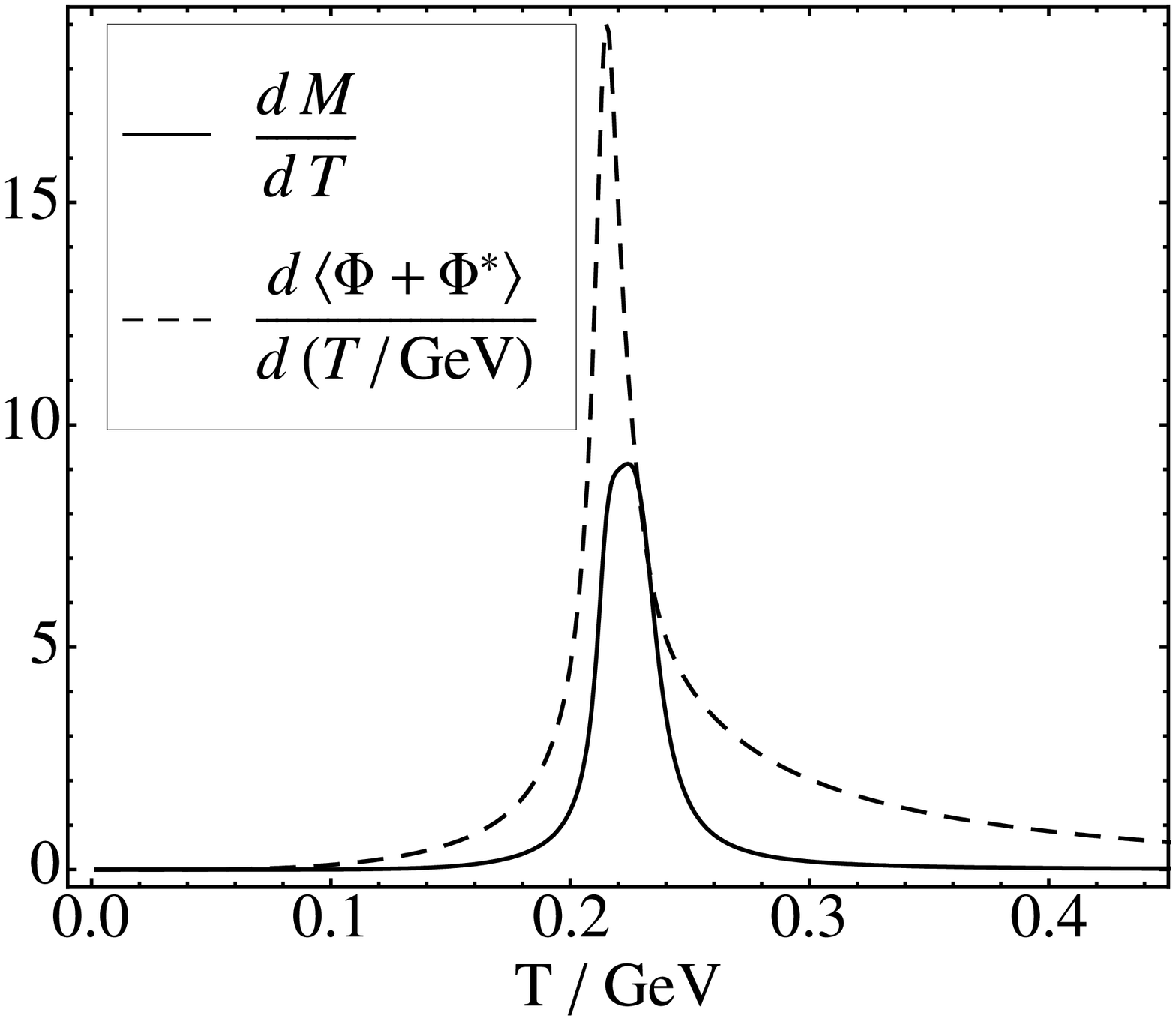}
\end{minipage}
\caption{
\label{fig:susceptiblity}
The chiral susceptibility $\chi_M$ (left panel solid line) and the Polyakov loop susceptibilities $\chi_{\Re \Phi}$ (left panel dashed line) and $\chi_{\Phi}$ (left panel dotted line) plotted as functions of temperature at vanishing quark chemical potential. These susceptibilities defined by Eqs.~(\ref{eqn:chi_m},\,\ref{eqn:chi_phi},\,\ref{eqn:chi_rephi}) and evaluated using Eq.~(\ref{eqn:gausssus}) are compared here to the derivative of the constituent quark mass (right panel solid line) and the expectation value of the real part of the Polyakov loop (right panel dashed line) with respect to temperature.}
\end{figure}

In Fig.~\ref{fig:susceptiblity} we show the chiral and the Polyakov loop susceptibilities as functions of temperature at vanishing quark chemical potential (left panel) and compare them to the temperature derivatives of the constituent quark mass and the Polyakov loop (right panel).
If we consider the behaviour of $\chi_M$, $\chi_{\Re \Phi}$ and $\chi_{\Phi}$ at $T\to 0$ we find that $\chi_{\Re \Phi}$ is finite, while $\chi_M$ and $\chi_{\Phi}$ vanish. 
This can be explained by the fact that $(\Re\Phi)^2 = \frac14(\Phi^2 + 2 \left| \Phi \right|^2 + {\Phi^*}^2)$ contains a $\mathrm{U}(1)$-symmetric term $\left| \Phi \right|$. As the $\mathrm{U}(1)$ symmetry incorporates the $Z(3)$ centre of $\mathrm{SU}(3)_\mathrm{c}$ this term does not have to vanish once the $Z(3)$ symmetry is fully restored at $T=0$.\footnote{The authors thank Chihiro Sasaki for pointing this out to them.}
The width of the peak in the temperature derivative of the dynamical quark mass $m = m_0-\sigma$ suggests that this crossover is influenced by the crossover of the Polyakov loop.
At finite current quark mass $m_0$ the PNJL model produces an approximate coincidence of the peaks in the susceptibilities of the Polyakov loop and the constituent quark mass $m$, consistent with the pattern observed in Fig.~\ref{fig:loopwithquarks}. 

\begin{figure}\centering
\includegraphics[width=.5\textwidth]{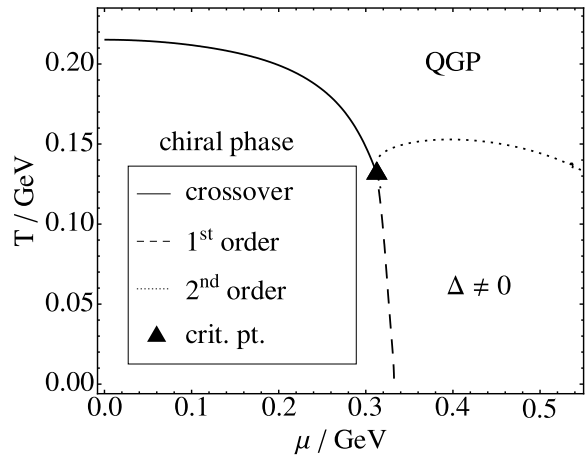}
\caption{
\label{fig:mfcomp}
Phase diagram implementing corrections to the order $\beta \leq 1$. Solid lines: crossover transition of the susceptibility related to the real part of the Polyakov loop, dashed lines: first order phase transition, and dotted: second order phase transitions.}
\end{figure}

A comparison of the phase diagram obtained in mean field approximation in Ref.~\cite{Roessner:2006xn} and the phase diagram including corrections to the order $\beta \leq 1$ shown in Fig.~\ref{fig:mfcomp}, explicitly approves that corrections to the phase diagram due to the fermion sign problem are indeed small \cite{Roessner:2006xn}: the influence of $\Im \Omega_0$ and the splitting of $\braket{\Phi^*}$ and $\braket{\Phi}$ are rather modest. 

\subsection{Moments of the pressure}

One benchmark for the PNJL model is its surprising capability of reproducing the trends of lattice QCD calculations.\footnote{Note however the discussion concerning the dependence on quark masses in Ref.~\cite{Ratti:2007jf}.} 
One way to handle the fermion sign problem in lattice QCD is to expand the calculated pressure about $\mu = 0$ in a Taylor series.
Such an expansion is given in Ref.~\cite{Allton:2005gk}:
\begin{align}
\label{eqn:pexpansion}
\frac{p(T,\mu)}{T^4}&=\sum_{n=0}^{\infty}c_n(T)
\left(\frac{\mu}{T}\right)^n & & \text{with } &c_n(T)
&=\left.\frac{1}{n!}\frac{\partial^n (p(T,\mu)/T^4)}{\partial(\mu/T)^n}\right |_{\mu=0}
\end{align}
with even $n$ as the situation is charge conjugation invariant. Specifically:
\begin{align}
 & \;c_2 = \left.\frac{1}{2}\,\frac{\partial^2 (p/T^4)}{\partial(\mu/T)^2}\right|_{\mu=0}, &
 & \;\,c_4 = \left.\frac{1}{24}\,\frac{\partial^4 (p/T^4)}{\partial(\mu/T)^4}\right|_{\mu=0}, \nonumber\\
 & c_6 = \left.\frac{1}{720}\,\frac{\partial^6 (p/T^4)}{\partial(\mu/T)^6}\right|_{\mu=0}, &
 & c_8 = \left.\frac{1}{40320}\,\frac{\partial^8 (p/T^4)}{\partial(\mu/T)^8}\right|_{\mu=0}.
\end{align}
The pressure in the PNJL model is evaluated by subtracting the divergent vacuum contributions of the thermodynamic potential:
\begin{equation}
\label{eqn:pressure}
p = -\left(\Omega - \Omega(T=0)\right)
\end{equation}

\begin{figure}
\begin{minipage}{.48\textwidth}
\includegraphics[width=\textwidth]{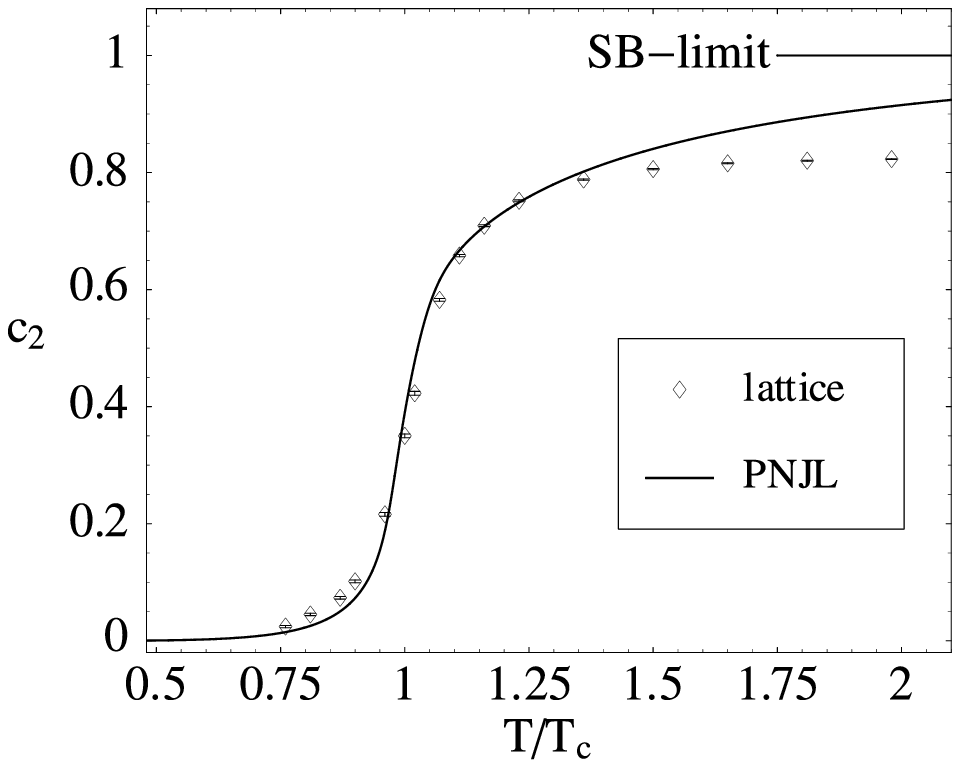}
\end{minipage}\hfill
\begin{minipage}{.485\textwidth}
\includegraphics[width=\textwidth]{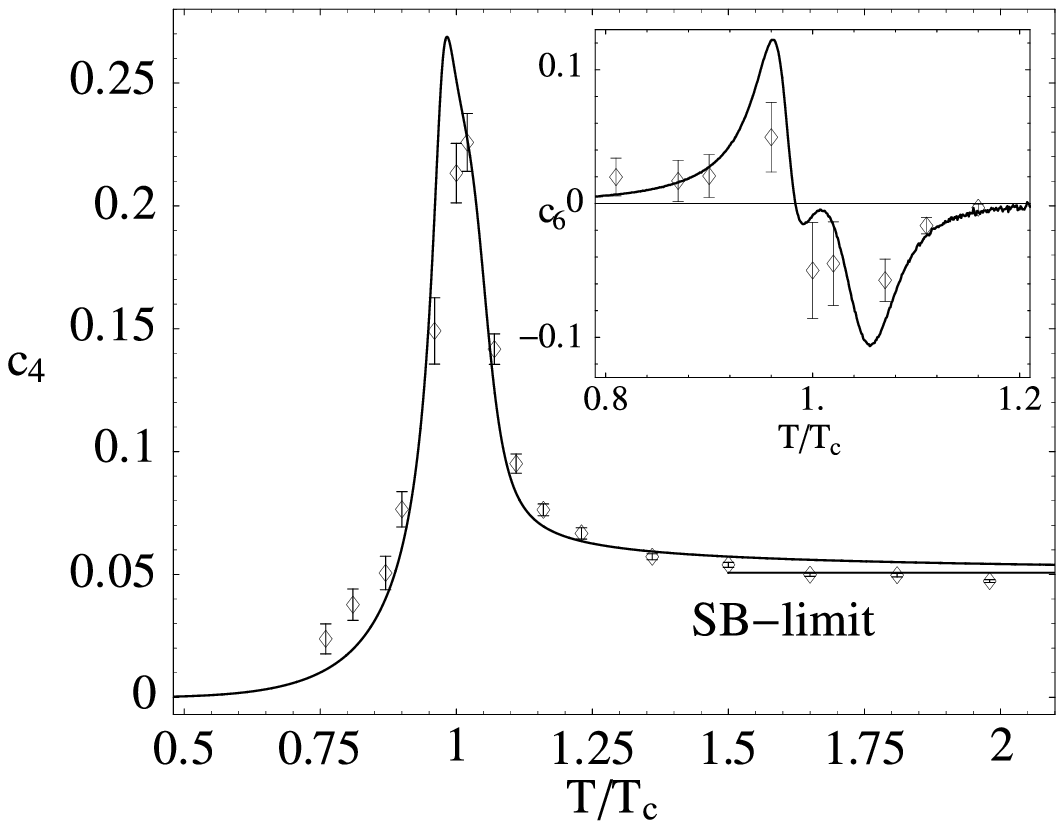}
\end{minipage}
\caption{\label{fig:moments}
The moments of the pressure with respect to $\frac{\mu}{T}$ as defined in Eq.~(\ref{eqn:pexpansion}). $c_2$ is shown in the left panel, $c_4$ is displayed to the right where $c_6$ is shown in the inset. The data deduced from lattice computations are taken from \cite{Allton:2005gk}. }
\end{figure}

Results for $c_2$, $c_4$ and $c_6$ are shown in Fig.~\ref{fig:moments}. 
In comparison with the plots for $c_n$ presented in a previous paper \cite{Roessner:2006xn} at the mean field level the moments $c_n$ show slightly more structure. The rise in $c_2$ is somewhat sharper, the peak in $c_4$ is about $5\,\%$ higher. 
In summary, however, the corrections induced so far by corrections involving $\braket{\Phi^*-\Phi}$ around the mean fields are small. 
Pionic fluctuations, to be discussed in Sec.~\ref{sec:pions}, tend to be more important. 
In presently available lattice results \cite{Allton:2005gk}, these latter effects are however suppressed by the relatively large pion masses. 

The ratio of the moments $c_4$ and $c_2$ has been discussed \cite{Ejiri:2005wq} as a suitable indicator of fluctuations close to the phase transition. 
The quantity of interest here is the cumulant ratio $R^{\mathrm{q}}_{4,2}$ defined in \cite{Ejiri:2005wq} and given as $R^{\mathrm{q}}_{4,2} = 12\,c_4/c_2$. 
The PNJL model calculation for this ratio is shown in Fig.~\ref{fig:rq42}. 
The dashed curve is found in the mean field limit with $\braket{\Phi^*}=\braket{\Phi}$ which suppresses one of the two Polyakov loop degrees of freedom. 
The solid curve is computed with inclusion of corrections beyond mean field and demonstrates the role of the non-zero $\braket{\Phi^*-\Phi}$. 
At temperatures below $T_{\mathrm{c}}$ one reaches $R^{\mathrm{q}}_{4,2} = 9$, the value characteristic of a hadronic resonance gas \cite{Ejiri:2005wq}.

\subsection{Interaction measure}

\begin{figure}
\begin{minipage}[t]{.485\textwidth}
\includegraphics[width=\textwidth]{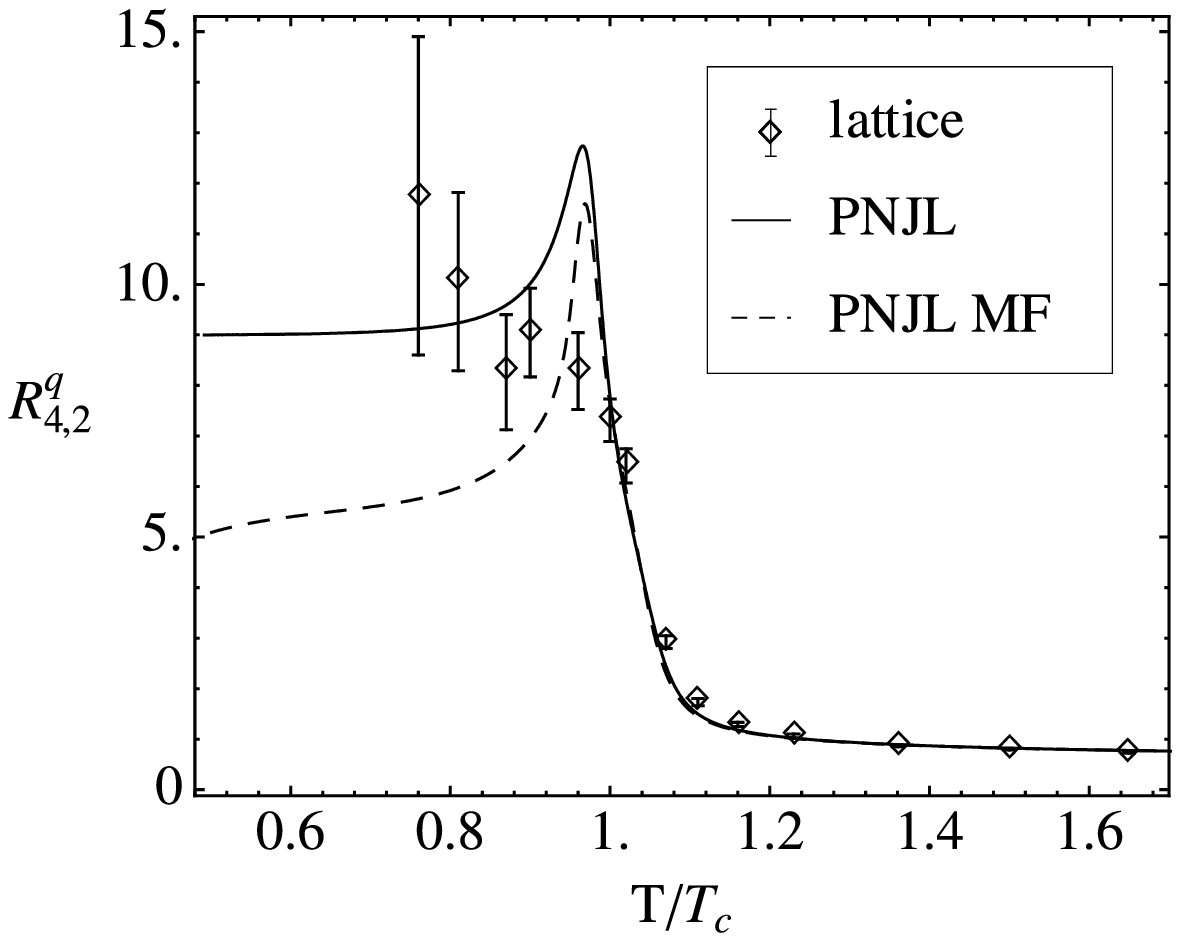}
\caption{\label{fig:rq42}
The cumulant ratio $R^{\mathrm{q}}_{4,2}$ from the PNJL model in and beyond mean field approximation in comparison with lattice QCD results, with $c_2$ and $c_4$ as given in Ref.~\cite{Allton:2005gk}. }
\end{minipage}\hfill
\begin{minipage}[t]{.48\textwidth}
\includegraphics[width=\textwidth]{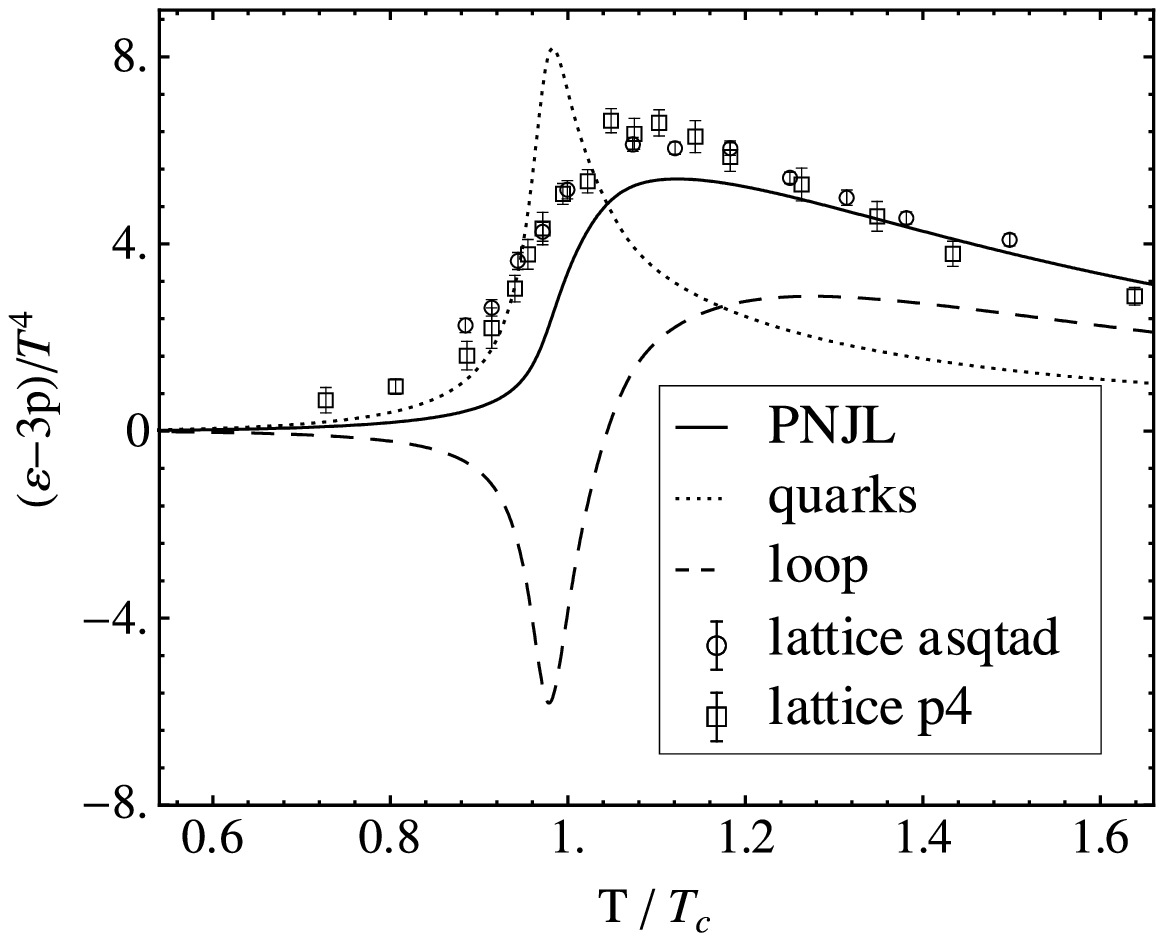}
\caption{\label{fig:iameasure}
Contributions to the conformal measure from quarks and Polyakov loop, as well as the total PNJL interaction measure for $N_{\mathrm{f}} = 2$. The $N_{\mathrm{f}} = 2+1$ lattice QCD results \cite{Karsch:2008fe} (with $N_\tau=8$ for p4-improved and asqtad action) are shown for orientation. }
\end{minipage}
\end{figure}

The PNJL results for the interaction or conformal measure $\varepsilon -3p$ are illustrated in Fig.~\ref{fig:iameasure}. 
The total interaction measure normalised to $T^4$ is split into quark and Polyakov loop parts. 
Note the sensitive balance between quark quasiparticle and Polyakov loop contributions to $\varepsilon -3p$ close to $T_{\mathrm{c}}$. 
In pure gauge QCD (or with infinitely heavy quarks) the Polyakov loop interaction measure is positive throughout. 
The presence of light quarks and their dynamical coupling to the Polyakov loop changes their pattern significantly. 
The Polyakov loop parts of the presure itself that determines the dashed curve in Fig.~\ref{fig:iameasure}, is found to be consistent with calculations reported in Ref.~\cite{Fukushima:2008wg}. 
For orientation, the total PNJL interaction measure (with $N_{\mathrm{f}}=2$) is shown in Fig.~\ref{fig:iameasure} along with recent $N_{\mathrm{f}}=2+1$ lattice QCD results \cite{Karsch:2008fe}.

\subsection{Speed of sound}

\begin{figure}
\begin{minipage}{.48\textwidth}
\includegraphics[width=\textwidth]{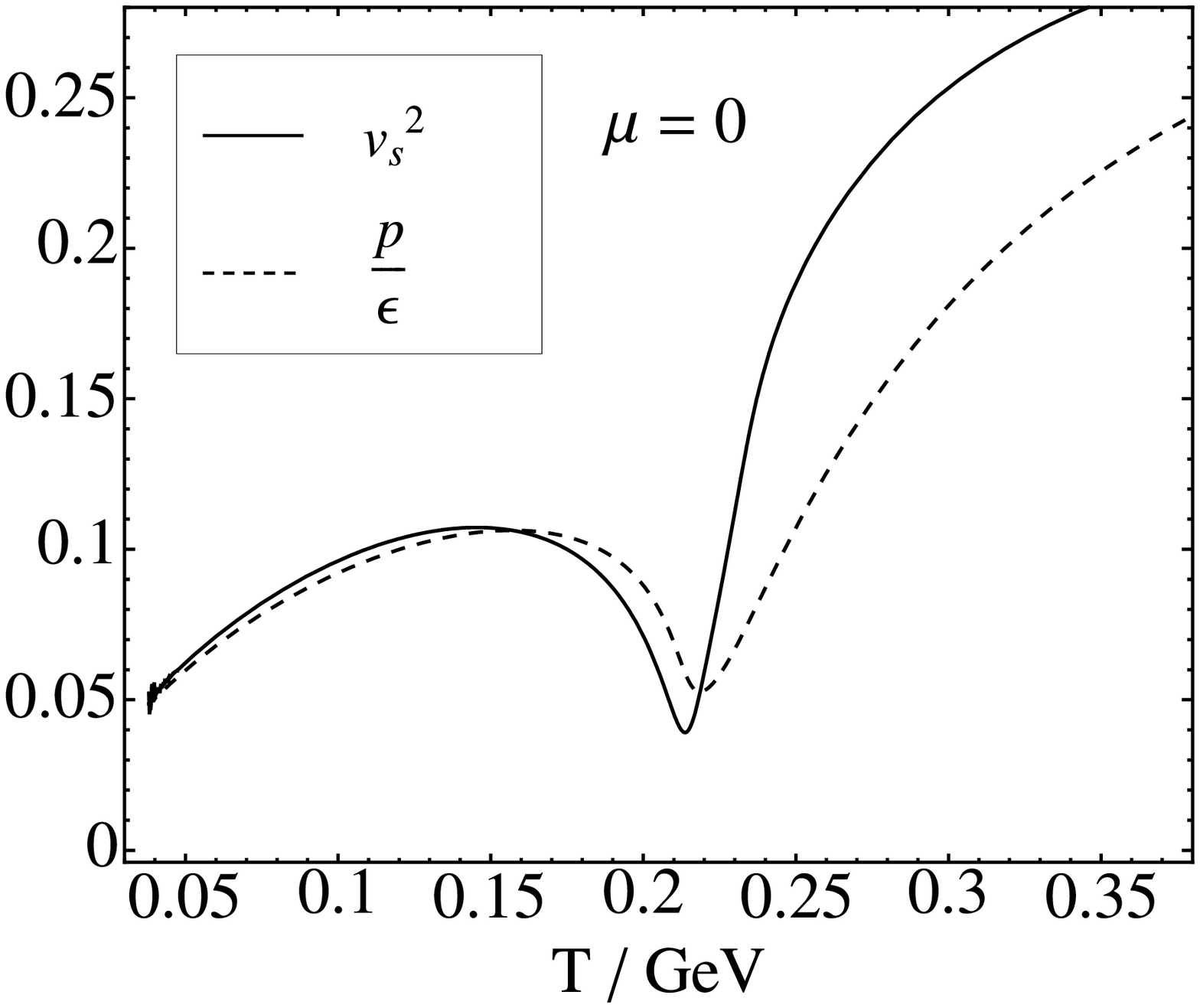}
\end{minipage}\hfill
\begin{minipage}{.485\textwidth}
\includegraphics[width=\textwidth]{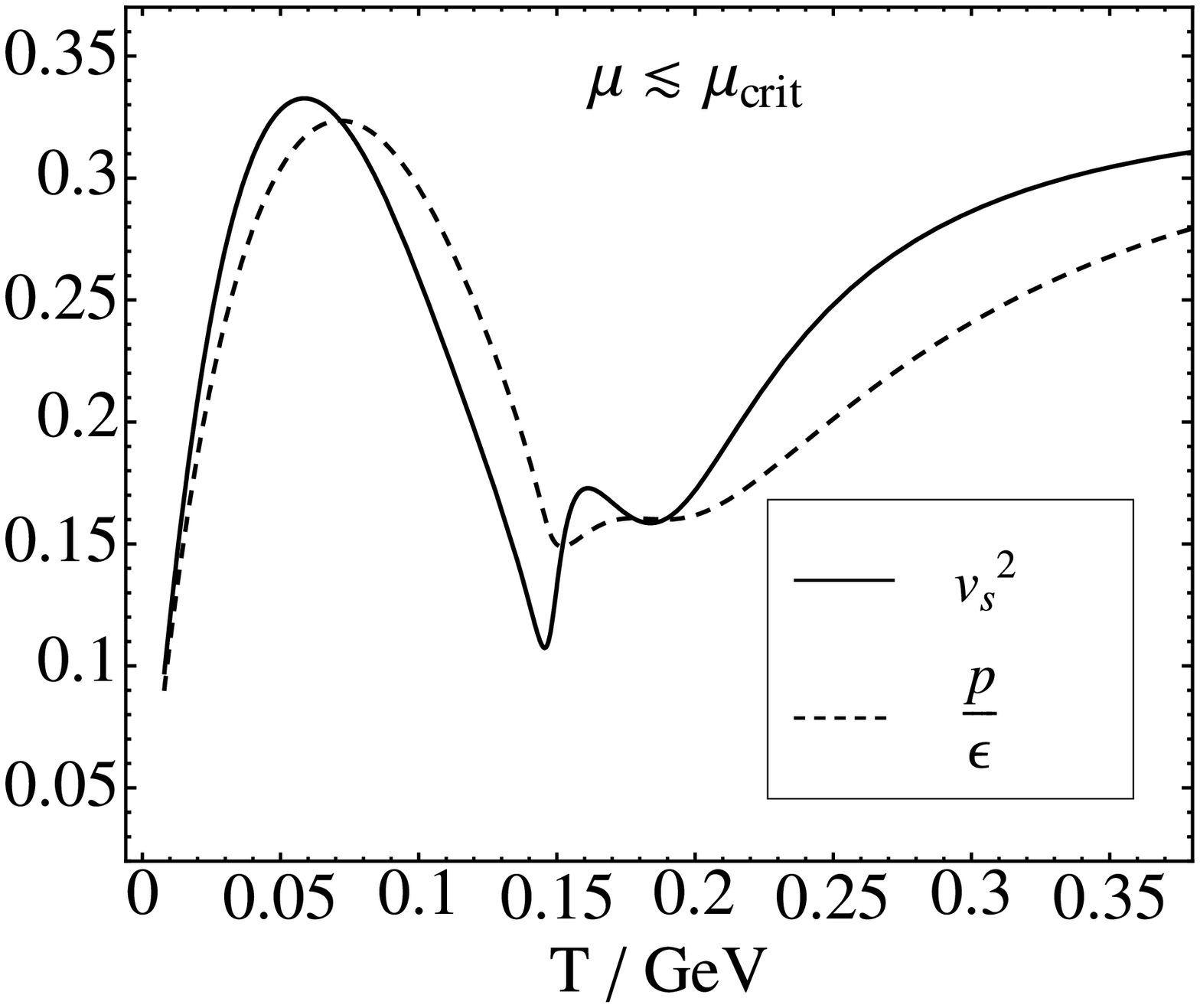}
\end{minipage}
\caption{
\label{fig:speed} The speed of sound (solid) and the ratio of pressure over energy density (dashed) at vanishing chemical potential as function of temperature (left panel). The right panel shows the same quantities at a quark chemical potential slightly less than the one at the critical point ($\mu  = 0.3\gev\lesssim \mu_\mathrm{crit} \simeq 0.31\gev$).
} 
\end{figure}

In Fig.~\ref{fig:speed} the squared speed of sound in units of the speed of light is plotted as a solid line. 
The speed of sound $v_\mathrm{s}$ is defined by 
\begin{equation}
v_\mathrm{s}^2 = \left. \frac{\partial p}{\partial \varepsilon} \right\vert_{S} = \left. \frac{\partial \Omega}{\partial T}\right\vert_{V} \bigg/ \left. T \frac{\partial^2 \Omega}{\partial T^2}\right\vert_{V} ~,
\end{equation}
where the denominator is the specific heat capacity $c_V$. 
The dashed line in Fig.~\ref{fig:speed} gives the size of the ratio of pressure and energy density, $\frac{p}{\varepsilon}$.
In the panel to the left where the quantities are plotted at vanishing chemical potential $\mu=0$ both graphs show a pronounced dip near the crossover transition temperature. 
In the panel to the right the same situation is plotted at a quark chemical potential close to the chemical potential of the critical point $\mu\lesssim \mu_{\mathrm{crit.}}$.

\section{Dynamical fluctuations}
\label{sec:pions}

So far the formalism presented has been focused on the treatment of fluctuations around the mean fields, averaged over space and (Euclidean) time. 
The homogeneous, constant Polyakov loop field and its corrections beyond mean field fall in this category. 

In this section we consider mesonic excitations and their propagation (i.\,e.\ $\frac{1}{N_{\mathrm{c}}}$-cor\-rec\-tions). The NJL framework is well suited to incorporate such effects. 
The NJL model features a dynamical mechanism which produces spontaneous chiral symmetry breaking and, at the same time, generates the pion as a Goldstone boson in the pseudoscalar quark-antiquark channel, together with a massive scalar (sigma) boson. 
The thermodynamics of these modes and their changing spectral properties have been subject of several NJL model calculations in the past \cite{Klevansky:1992qe,Hatsuda:1994pi,Lutz:1992dv}.

With increasing temperature, the mass of the pion is still protected by its Goldstone boson nature, whereas the sigma mass drops until at $T \sim T_\mathrm{c}$ it becomes degenerate with the pion, signalling restoration of chiral symmetry in its Wigner-Weyl realisation. 
For $T>T_\mathrm{c}$, the $\pi$ and $\sigma$ masses jointly increase quite rapidly while at the same time their widths for decay into $q\bar{q}$ grow continuously. 
This implies that at temperatures exceeding $T_\mathrm{c}$ both $\pi$ and $\sigma$ modes become thermodynamically irrelevant while correlated quark-antiquark pairs carrying the quantum numbers of $\pi$ and $\sigma$ can still be active above $T_\mathrm{c}$. 
One therefore expects that the corrections to the pressure from propagating pions and sigmas should be concentrated around $T_\mathrm{c}$. 
These mesonic modes are colour singlets\footnote{Colour octet quark-antiquark modes turn out to be heavy and far removed from the spectrum of active degrees of freedom.}. 
Thus their statistical weight is much smaller than the weight of the deconfined quark quasiparticles.

\subsection{Meson propagators in the PNJL model}

We start from the derivation of mesonic propagators in the PNJL model as performed, for example, in \cite{Hansen:2006ee}. 
We calculate the momentum dependent propagator
\begin{equation}
\label{eqn:mesonpropagator}
\parbox{40pt}{\begin{picture}(40,40)
\put(0,20){\includegraphics{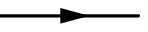}}
\put(2,10){$j$}\put(32,10){$k$}\put(17,30){$q^\mu$}
\end{picture}} = \left[  \frac{\partial^2 \mathcal{S}_{\mathrm{bos}}}{\partial \xi_j(q^\mu) \partial \xi_k(-q^\mu)}\right]^{-1}~,
\end{equation}
where $\xi = \theta -\theta_{\mathrm{MF}}$ now stands for the pion field or for the deviation of the sigma field from its expectation value. 
Note that the functional trace in the formula for $\mathcal{S}_{\mathrm{bos}}$ ensures momentum conservation, such that the sum of the momentum arguments in the denominator always vanishes.
The calculation can be done numerically as it was done in the previous section. 
Alternatively, we use an analytic approach as follows. 
Recall some useful formulae also exploited in Refs.~\cite{Allton:2003vx,Allton:2005gk}:
\begin{align}
\frac{\partial \ln \det M}{\partial x} &= \tr \left[ M^{-1}\frac{\partial M}{\partial x} \right] &\text{and}&& \frac{\partial M^{-1}}{\partial x} &= -M^{-1}\frac{\partial M}{\partial x}M^{-1}\;,
\end{align}
with $M$ an invertible matrix and $\frac{\partial M}{\partial x}$ is the component-wise derivative of this matrix. 
Applying this to the PNJL action $\mathcal{S}_{\mathrm{bos}}$ in (\ref{eqn:omegageneral}) and neglecting the potential terms for the moment we find 
\begin{equation}
\label{eqn:firstderiv}
\frac{\partial \mathcal{S}_{\mathrm{bos}}}{\partial \theta} = -\frac{V}{2}\sum_n\int\frac{\intd^3p}{\left(2\pi\right)^3}\Tr\left[ \tilde{S}\left(i\omega_n,\vec{p}\,;\theta\right) \frac{\partial \tilde{S}^{-1}\left(i\omega_n,\vec{p}\,;\theta\right)}{\partial \theta}\right]~,
\end{equation}
where $\tilde{S}^{-1}\left(i\omega_n,\vec{p}\,;\theta\right)$ denotes the inverse quark propagator with emphasis on the fact that the quark propagates in the mesonic background field $\theta$.
This formalism makes it possible to calculate derivatives with respect to bosonic fields (say $\theta_k$) that have not been explicitly included in the action, as long as it is ensured that the model does not produce finite vacuum expectation values for these particular fields.
Not having a vacuum expectation value is equivalent to the fact, that the mean field equations corresponding to these fields are satisfied for a vanishing field, i.\,e.\ that 
\begin{equation}
\left.\frac{\partial \mathcal{S}_{\mathrm{bos}}}{\partial \theta}\right\vert_{\theta_k = 0} = 0\;.
\label{eqn:mfthetak}
\end{equation}
All we need to know is the constant matrix $\frac{\partial \tilde{S}^{-1}}{\partial \theta_k}$.
This matrix involves the Dirac, colour and flavour structure of a quark-antiquark pair (or a quark-quark pair) that couples to the bosonic field $\theta_k$, i.\,e.\ it is determined by the quantum numbers of $\theta_k$.
The mean field equation is fulfilled if the trace in Eq.~(\ref{eqn:firstderiv}) vanishes for the given Dirac, colour and flavour structure.
For the pion field this is true as long as there is no pion condensate.\footnote{The mean field equation is satisfied as the flavour-trace $\tr_{\mathrm{f}}[\,\Eins \,\tau_i\,]=0$ with $i=1,2,3$  vanishes.}
The condensate corresponding to the sigma, namely the chiral condensate, figures explicitly in the action and is therefore included in the quark propagator. 

In the case of the pion and sigma propagators the functional derivative in (\ref{eqn:mesonpropagator}) produces exactly the trace over Dirac, colour and flavour structures known from RPA calculations \cite{Klevansky:1992qe,Hansen:2006ee,Hatsuda:1994pi,Lutz:1992dv}.
We adopt the definition of the quark distribution functions $f^+_\Phi $ and $f^-_\Phi $ and the separation of the emerging integral into the contributions $I_1$ and $I_2$ as given in Ref.~\cite{Hansen:2006ee}. 
In the treatment of the thermodynamics we have modified the cutoff prescription of the standard NJL model, such that non-divergent integrals are integrated over the whole quark-momentum range, while only divergent integrals are regularised by the usual NJL three-momentum cutoff. 
The separation of finite and divergent contributions is defined such that the model reproduces the classical limit at high temperatures, i.\,e.\ the Stefan-Boltzmann limit.
As a downside, for consistency all newly appearing integrals have to be treated in the same manner, which leads to slightly different results from those given in Ref.~\cite{Hansen:2006ee}.

\subsection{Mesonic corrections to the pressure}

Once the meson propagators are given, it is possible to evaluate the contribution to the pressure from mesons propagating in the heat bath using RPA methods. 
Applying Bethe-Salpeter (RPA) equations generates spectral functions 
$$\rho_{\mathrm{M}}(\omega,\,\vec{q};\,T)%= \frac{1}{G}\Im\,\tilde{S}_\mathrm{M}(\omega,\,\vert\vec{q}\,\vert=0)
 = \frac{G \Im \Pi_\mathrm{M}(\omega,\,\vec{q};\,T)}{(1-G \Re  \Pi_\mathrm{M})^2+ (G \Im \Pi_\mathrm{M} )^2} $$ 
with the thermal quark-antiquark polarisation function
\begin{equation*}
% \label{eqn:polarization}
\Pi_{\mathrm{M}}(\omega,\vec{q};\,T) = T \sum_{\omega_n} \int\frac{\intd^3 p}{(2\pi)^3} \Tr\left[ \Gamma_\mathrm{M} \tilde{S}(\iu\omega_n+\mu,\vec{p}\,) \Gamma_\mathrm{M} \tilde{S}(\iu(\omega_n-\omega)+\mu,\vec{p}-\vec{q}\,) \right]~,
\end{equation*}
where the sum is taken over the Matsubara frequencies $\omega_n = (2n+1)\pi\, T$. 
Here $\Gamma_\mathrm{M}$ is a Dirac, flavour and colour representation of a meson current labelled $\mathrm{M}$. 
In this work we only focus on the pseudoscalar isovector channel (i.\,e.\ pionic excitations) and the scalar isoscalar channel. 
$\tilde{S}(\iu\omega_n,\vec{p}\,) = -\frac{m+\slashed{p}}{\omega_n^2+p^2+m^2}$ denotes the quark quasiparticle propagator with $\slashed{p}=\iu\omega_n\gamma_0-\vec{\gamma}\cdot\vec{p}$. 

\begin{figure}
\begin{minipage}{.48\textwidth}
\includegraphics[width=\textwidth]{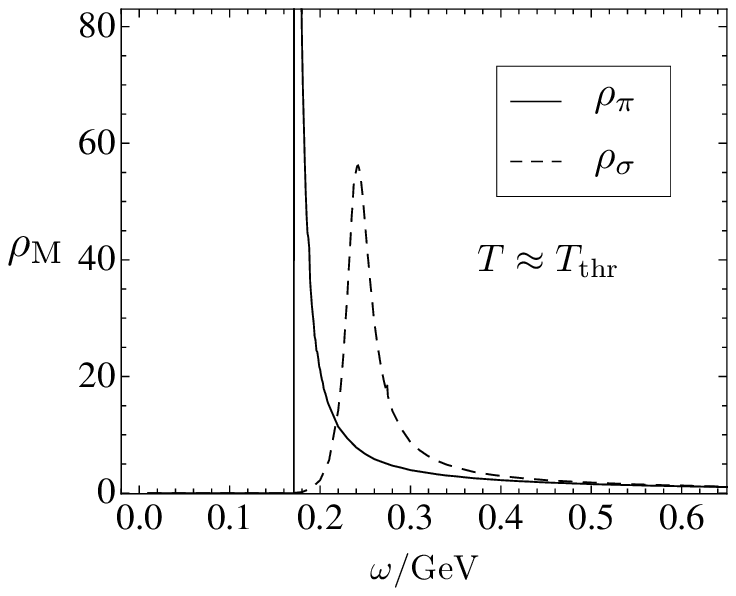}
\end{minipage}\hfill
\begin{minipage}{.48\textwidth}
\includegraphics[width=\textwidth]{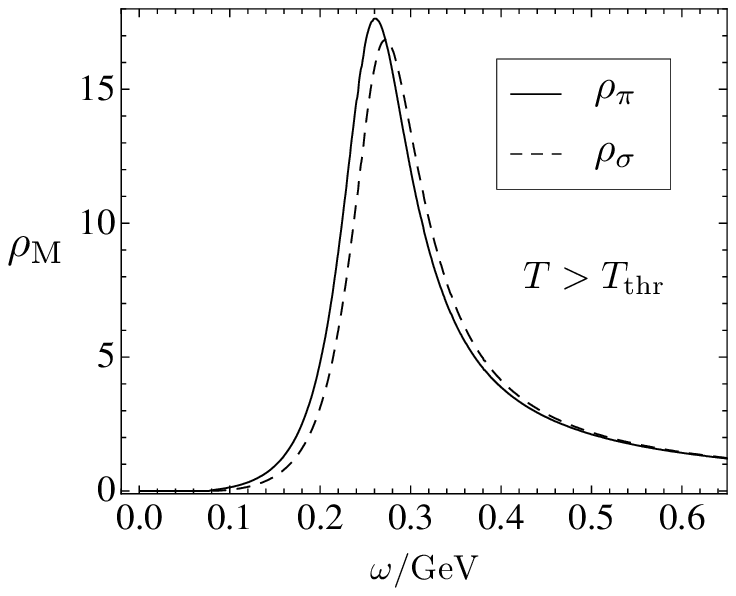}
\end{minipage}
\caption{The spectral functions $\rho_{\mathrm{M}}= \frac{G \Im \Pi_\mathrm{M}}{(1-G \Re \Pi_\mathrm{M})^2+ (G \Im \Pi_\mathrm{M} )^2} $ taken at $\vec{q}=0$ for pion and sigma at $T\approx T_{\mathrm{thr}} $ (left) and at $T> T_{\mathrm{thr}} $ (right). \label{fig:spectral}}
\end{figure}

The pressure below $T_{\mathrm{c}}$ is essentially generated by the pion pole with its almost temperature independent position. 
Therefore the calculated pressure below $T_{\mathrm{c}}$ basically represents the one of a pion gas with fixed (temperature independent) mass. 
Fig.~\ref{fig:spectral} shows, as examples, the spectral functions for the pion and sigma modes at threshold temperature $T_{\mathrm{thr}}$ where the breakup into a quark-antiquark pair occurs. 
This threshold temperature is at about $ 1.1 \,T_{\mathrm{c}}$. 
At this point the $\pi$ and $\sigma$ spectral functions are still distinguishable (left panel of Fig.~\ref{fig:spectral}), whereas they coincide (right panel) at temperatures well above threshold where $\pi$-$\sigma$ degeneracy indicates restoration of chiral symmetry in its Wigner-Weyl realisation. 
Their width is a measure of the decay of the (increasingly massive) pionic and sigma modes into (light) deconfined quark-antiquark pairs at temperatures above $T_{\mathrm{c}}$. 

The resonant interaction of instable mesons with the quark sea above $T_{\mathrm{c}}$ produces an additional pressure contribution.
This contribution is not part of the quark pressure previously calculated in Hartree-Fock approximation. 
The meson decay products form rings of RPA chains. 
Such kind of pressure contributions are investigated in Ref.~\cite{Hufner:1994ma} and calculated performing the ring sum. 
However, below $T_{\mathrm{c}}$ the NJL model does not handle the mesonic degrees of freedom  properly. 
In the hadronic phase the coupling of mesonic modes to the quark-antiquark continuum is suppressed by confinement, whereas $\rho_{\mathrm{M}}$ receives contributions from decays into $q\bar{q}$ even below $T_{\mathrm{c}}$. 
This unphysical feature persists \cite{Hansen:2006ee} in the PNJL generalisation of the NJL approach. 
Moreover, the non-renormalisability of the NJL model requires to introduce further subtractions when following the lines of Ref.~\cite{Hufner:1994ma}. 
To avoid such arbitrariness and unphysical features we ignore the decay of meson modes into $q\bar{q}$-pairs altogether when calculating an estimate for the meson contributions to the pressure: 
\begin{equation} % \;\Theta(2m_\mathrm{q}-m_{\mathrm{pole}})
\label{eqn:omegacorr}
\delta\Omega =\nu %\,\bigg\lbrace 
\int\frac{\intd^3 q}{(2\pi)^3} \,T \ln(1-\ee^{-{E_q}/{T}}) + B(T) ~,
\end{equation}
where $\nu$ is the statistical weight of the corresponding meson species, $E_q = \sqrt{{\vec{q}\,}^2+m^2_{\mathrm{pole}}(T)}$ with $m_{\mathrm{pole}}(T)$ the temperature dependent pion and sigma pole mass determined by $1-G\Re \Pi = 0$. Furthermore $B(T)$ is an appropriately chosen vacuum energy constant ensuring thermodynamic consistency. $B(T)$ is fixed such that the temperature dependence of the pole mass $m_{\mathrm{pole}}(T)$ is compensated on differentiating $\Omega$ with respect to the temperature $T$. 
This implies that the inclusion of $B(T)$ ensures that $\left.\partial \Omega/\partial m_{\mathrm{pole}}\right\vert_T =0$.

\begin{figure}
\begin{minipage}{.48\textwidth}
\includegraphics[height=.76\textwidth]{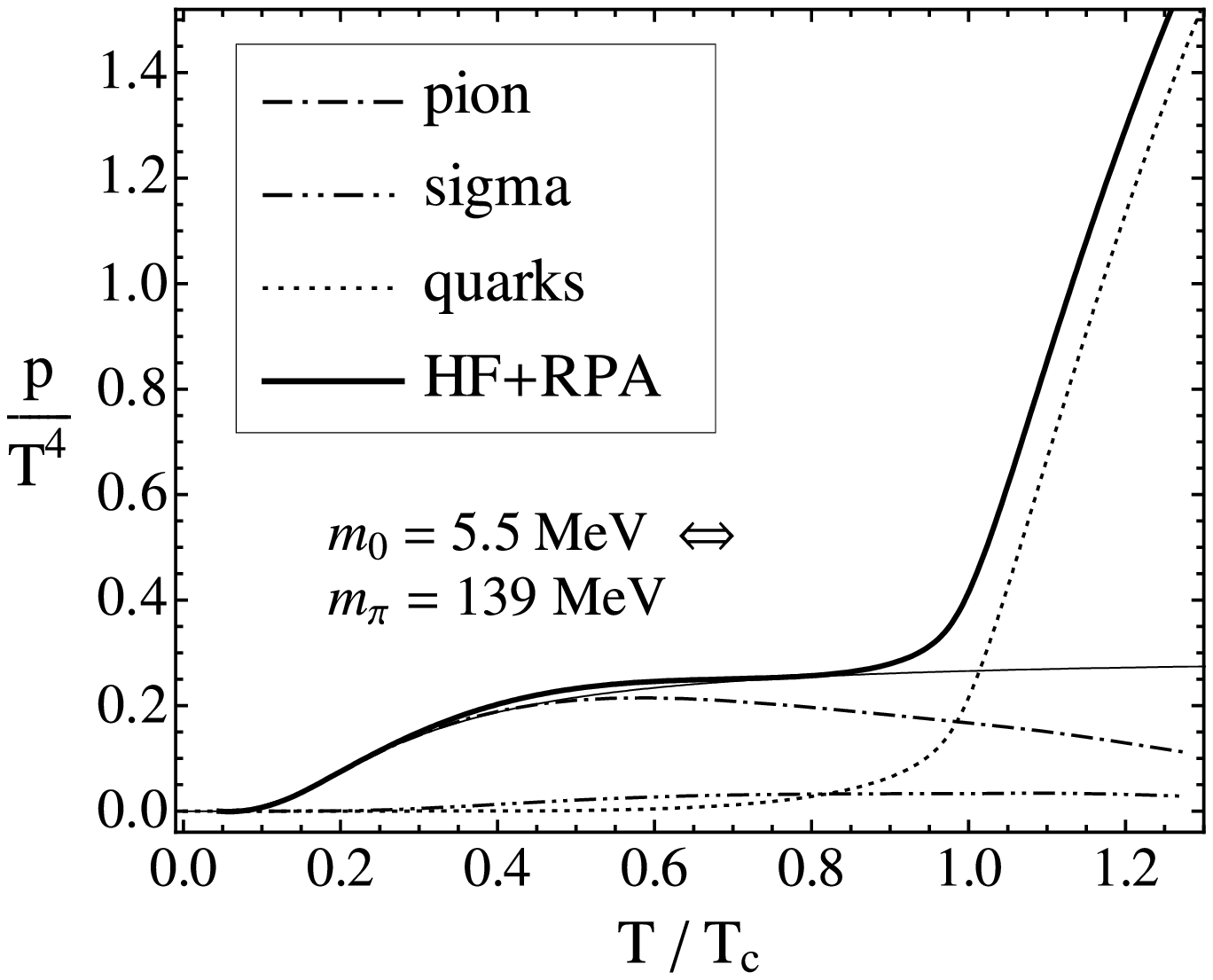}
\caption{The pressure contribution originating from pion modes, sigma modes and from quarks in Hartree-Fock approximation (dotted). The thin solid line represents the pressure of a gas of bosons with three internal degrees of freedom and a constant mass $m=m_\pi(T=0)$.  \label{fig:totalpressure}}
\end{minipage}\hfill
\begin{minipage}{.48\textwidth}
\includegraphics[height=.76\textwidth]{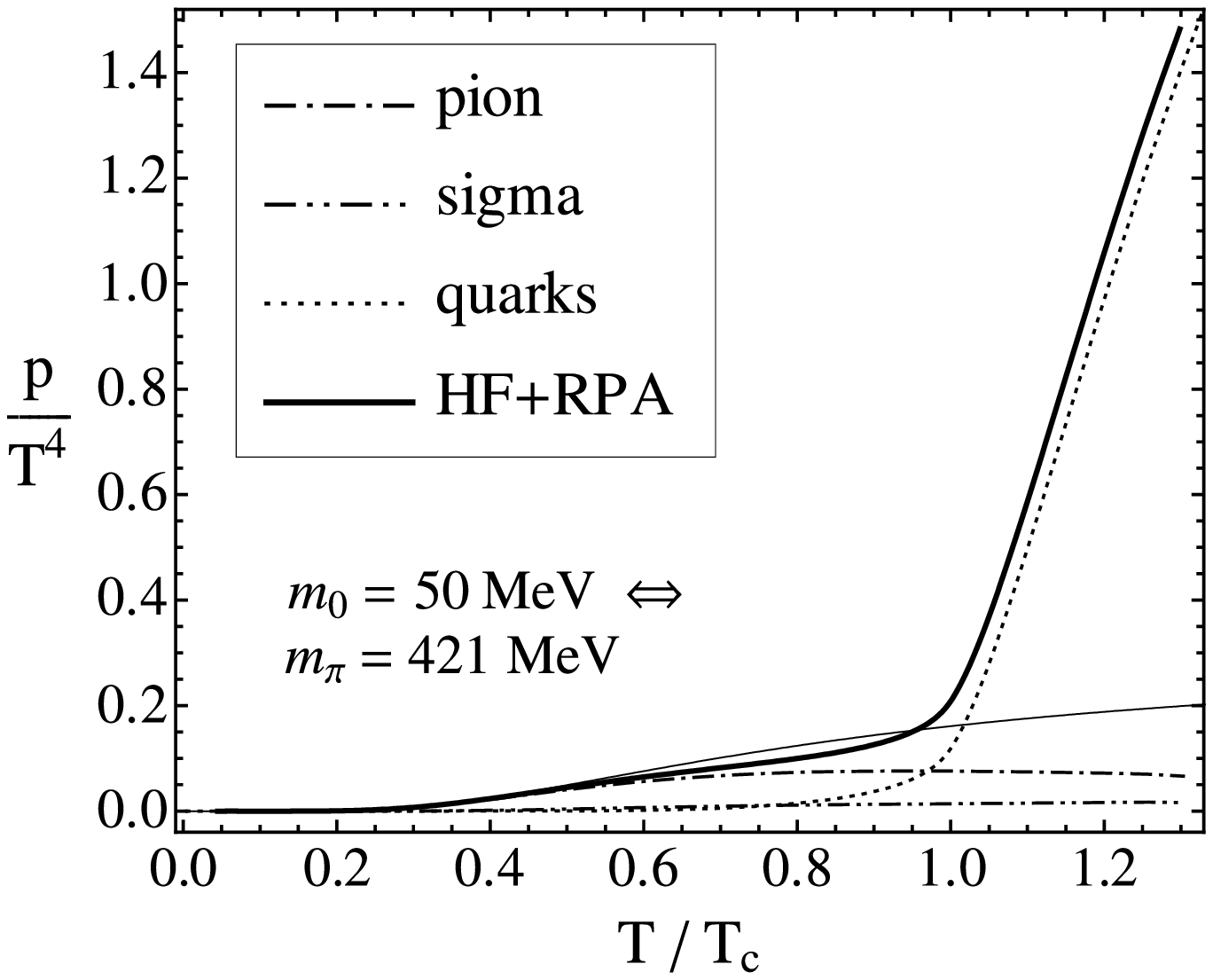}
\caption{Same as Fig.~\ref{fig:totalpressure}, but with higher current quark mass $m_0=50\mev \Rightarrow m_\pi=421\mev$ (compared to $m_0=5.5\mev \Rightarrow m_\pi=139\mev$ in Fig.~\ref{fig:totalpressure}). The pressure of the boson gas (thin solid line) was now plotted using the heavier pion mass. \label{fig:totalpressure50}\vspace{14pt}}
\end{minipage}
\end{figure}

In Fig.~\ref{fig:totalpressure} the calculated pressure of $\pi^{0,\pm}$ and sigma modes are compared with the quark Hartree-Fock pressure and the result for the overall pressure of Hartree-Fock plus RPA is plotted. 
For comparison the pressure of a Bose gas with three internal degrees of freedom is indicated by the thin solid line. 
Below the crossover temperature $T_\mathrm{c}$ one can clearly identify the pion gas contribution resulting from the RPA calculations. 
Once the meson masses reach the scale of the NJL cutoff $\Lambda$ the used approximation breaks down. 
The inversion of the scale hierarchy appears at temperatures of about $1.3\,T_{\mathrm{c}}$. 

For larger current quark masses the meson gas contributions and correlations are reduced. 
This effect is illustrated by Fig.~\ref{fig:totalpressure50} where the pressure of the PNJL model is plotted using an increased current quark mass leading to an unphysically heavy pion. 
Thus for heavy pions the agreement with lattice data observed in a previous publication \cite{Roessner:2006xn} remains. 
This agreement is also confirmed by calculations in a non-local PNJL framework \cite{Blaschke:2007np} which does not suffer from cutoff artefacts. 

\begin{figure}
\centering
\includegraphics[height=.46\textwidth]{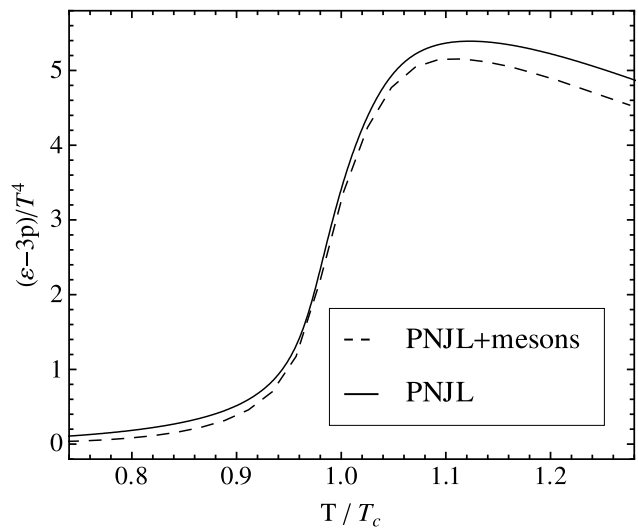}
\caption{
The normalised interaction measure $(\varepsilon-3p)/T^4$ from the PNJL model with and without mesonic corrections. 
% The mesonic corrections to the interaction measure are small. 
\label{fig:mesoniameasure}}
\end{figure}

The mesonic contribution to the interaction measure is rather modest. 
The interaction measure already shown in Fig.~\ref{fig:iameasure} is replotted in Fig.~\ref{fig:mesoniameasure} including mesonic contributions. 

\section{Conclusions and outlook}
\label{sec:conclusio}

The PNJL model as an approximation to QCD thermodynamics picks up on two basic properties of low-energy QCD: spontaneous chiral symmetry breaking and confinement.
In this work some of the existing calculations \cite{Ratti:2005jh,Ratti:2006wg,Roessner:2006xn} have been extended in several directions.
We have reviewed the expectation values of the Polyakov loop and its complex conjugate, the phase diagram, the moments of the pressure and the speed of sound in a framework beyond mean field theory.
While the phase diagram does not show significant changes when improving the mean field approximation, the moments of the pressure and the speed of sound show quantitative differences on the order of $5\,\%$. In general the structures observed become more articulate.
In the case of the Polyakov loop and its complex conjugate the corrections cause qualitative differences.
While the Polyakov loop and its complex conjugate are equal at mean field level in the present approach, the corrections beyond mean field generate the split of the two expectation values $\braket{\Phi}$ and $\braket{\Phi^*}$ at non-zero quark chemical potential.
The numerical results show that the corrections are largest in the vicinity of phase transitions or rapid crossovers. 
This comes as no surprise as it is the transitional region between two phases where we expect large fluctuations. 

The degrees of freedom that govern the low temperature regime, primarily the pions, produce significant corrections to the pressure only in the regime below the critical temperature $T_{\mathrm{c}}$ where constituent quarks are frozen and confined. 
As soon as the pressure of quark degrees of freedom starts to rise at the chiral and deconfinement crossover, mesonic pressure contributions become comparatively small. 

The good agreement of the PNJL model at mean field level with lattice calculations remains in the presence of the mesonic corrections calculated in this work. 
The most prominent feature of the pressure, namely the steep rise near the critical temperature, is only slightly modified by the corrections due to dynamic fluctuations of pions and sigma mesons. 
Below the quark-antiquark threshold the pressure generated by pion fluctuations is basically the pressure of a free pion gas. 
As low temperatures are difficult to access by lattice QCD, the pressure in typical lattice calculations is usually normalised to zero at some finite temperature below $T_{\mathrm{c}}$. 
This might explain why the pressure of the PNJL model including mesonic corrections is slightly higher than the pressure resulting from lattice calculations \cite{Allton:2005gk}. 
Due to these normalisation issues the comparison suffers from this uncertainty, $\Delta (p/T^4) = \left. p/T^4\right\vert_{T=T_{\mathrm{norm}}}-\left.p/T^4\right\vert_{T\to 0}$, which in turn depends on the normalisation temperature $T_{\mathrm{norm}}$ and additionally on the realized pion mass. 
For large pion masses (see Fig.~\ref{fig:totalpressure50}) this correction is small maintaining the good agreement between PNJL and lattice results. 
For small pion masses the pressure contribution from pion modes is almost flat in the temperature region $T\approx m_\pi$, such that the pressures from lattice and PNJL calculations mainly differ by a shift in $p/T^4$. 
Shifting the lattice data to higher values of $p/T^4$ indeed reduces the difference between Stefan-Boltzmann limit and lattice data for the pressure at high temperatures around $2$--$3\, T_\mathrm{c}$ and above, improving the agreement between lattice results and PNJL. 
Even when taking into account these issues in the comparison of PNJL and lattice results, we conclude that there exists a good qualitative and quantitative agreement of these two approaches.

\subsection*{Acknowledgements}
We thank Marco Cristoforetti, Kenji Fukushima and Volker Koch for stimulating discussions.

\appendix
\label{app:corrections}
\section{Detailed derivation of corrections to mean fields}

\subsection{Expansion of the effective action}

This appendix displays some technical details concerning the treatment of fluctuation corrections  beyond mean field approximation in the PNJL model (cf. Sec.~\ref{sec:corr}). 

In the following we denote by $\theta = (\theta_i)$ the set of fields $(\sigma,\,\Delta,\,\phi_3,\,\phi_8)$ which operate as bosonic degrees of freedom in the effective action $\mathcal{S}_{\mathrm{bos}}$ of Eqs.~(\ref{eqn:omegageneral}) and (\ref{eqn:bosaction}). 
Furthermore, let $\theta_0 = (\braket{\sigma}_0,\,\braket{\Delta}_0,\,\braket{\phi_3}_0,\,\braket{\phi_8}_0)$ be the set of mean field (expectation) values of these quantities, and introduce deviations from the mean fields by $\xi = (\xi_i) = \theta-\theta_0$. 

A frequently used procedure that we follow here, is to expand the effective action in powers of $\xi$ around a properly chosen mean field configuration. 
The Gaussian part of such an expansion of the path integral can be handled analytically. 
In Sec.~\ref{sec:corr} the mean field approximation has been defined such that the (formally) complex action $\mathcal{S}_{\mathrm{bos}}$ produces, to this leading order, a real-valued thermodynamical potential (or pressure), $\Omega_{\mathrm{MF}} = \Rens\left[\Omega_0\right]$, subject to the mean field equations (\ref{eqn:mfeqn}). 
The expansion of $\mathcal{S}_{\mathrm{bos}}$ is then of the generic form
\begin{equation}
\label{eqn:expansion}
\mathcal{S}_{\mathrm{bos}} = \frac{V}{T}\left( \Omega_{\mathrm{MF}} + \omega^{(1)}\cdot\xi +\frac12 \xi\cdot\omega^{(2)}\cdot\xi \cdots \right)~,
\end{equation}
where we have introduced the notations $a\cdot b = \sum_i a_i\,b_i$ and $a\cdot A \cdot b = \sum_{ij} a_i\,A_{ij}\,b_j$, with summations extending over all bosonic degrees of freedom. 
The expansion (\ref{eqn:expansion}) is performed such that the path integral is optimally approximated. 
This is achieved when the perturbative terms in the expansion of the action are maximally suppressed. 
With the thermodynamic weight $\ee^{\mathcal{-S}}\in\mathds{C}$ this approximation is optimal near the maximum of $\left\vert \ee^{-\mathcal{S}} \right\vert$. 
The equations to determine $\theta_0$ are the mean field equations (\ref{eqn:mfeqn}) (also used in \cite{Roessner:2006xn}). 

Given the expansion (\ref{eqn:expansion}) in terms of the $\xi$ fields, thermal expectation values incorporate fluctuations around the mean field configuration $\theta_{\mathrm{MF}}\equiv\theta_0$. 
We refer to these corrections as ``fluctuations'' even if the fields themselves (such as the Polyakov loop field variables $\phi_3$ and $\phi_8$) are constant in space and time. 

A perturbative approach is now used to calculate corrections to the mean field solutions. 
The action $\mathcal{S}_{\mathrm{bos}}$ is split into ``large'' and ``small'' parts, $\mathcal{S}_{\mathrm{bos}} = \mathcal{S}_{0} + \mathcal{S}_{\mathrm{I}}$, as follows: the ``large'' part $\mathcal{S}_{0}$ incorporates the leading mean field terms plus the additional Gaussian part of $O(\xi^2)$ in Eq.~(\ref{eqn:expansion}):
\begin{equation}
\label{eqn:esnull}
 \mathcal{S}_{0}  = \frac{V}{T}\left(  \Rens\left[\Omega_0\right]+\frac12\xi\cdot\omega^{(2)}\cdot\xi  \right)~,
\end{equation}
while $\mathcal{S}_{\mathrm{I}}$ deals with the remaining pieces, in particular with the non-vanishing $\Imns\left[\Omega_0\right]$. 
The leading correction of this sort is the term $\delta \mathcal{S}_{\mathrm{I}} = \frac{V}{T}\omega^{(1)}\cdot\xi$. 
In the present context we truncate Eq.~(\ref{eqn:expansion}) as it stands and keep only this term in $\mathcal{S}_{\mathrm{I}}$, for the moment. 

The thermal expectation values of a given quantity $f(\xi)$ is proportional to 
\begin{equation}
\label{eqn:expvalue}
\int \mathcal{D}\xi\;f(\xi)\,\ee^{-\mathcal{S}_{\mathrm{bos}}} = \int \mathcal{D}\xi\;f(\xi)\,\ee^{-\mathcal{S}_{0}} \,\ee^{-\mathcal{S}_{\mathrm{I}}} ~, 
\end{equation}
where, for fields constant in space-time, the path integral reduces to
\begin{equation}
\label{eqn:expvaluexi}
\int \intd\xi\;f(\xi)\,\ee^{-\mathcal{S}_{0}(\xi)} \,\ee^{-\mathcal{S}_{\mathrm{I}}(\xi)} = \int \intd \xi\;f(\xi)\,\ee^{-\mathcal{S}_{0}(\xi)} \,\ee^{-\iu k\cdot \xi} ~, 
\end{equation}
with
\begin{equation}
 \label{eqn:kdef}
k= \frac{V}{\iu T} \omega^{(1)} = \frac{V}{T} \Im\omega^{(1)}~. 
\end{equation}
A perturbative expansion of $f(\xi)$ about $\xi=0$ (i.\,e.\ about $\theta = \theta_{\mathrm{MF}}$) in powers of $\xi$ involves integrals of the form
\begin{equation}
\label{eqn:fourrier}
\int \intd\xi\;\xi^n\,\ee^{-\mathcal{S}_{0}(\xi)} \,\ee^{-\iu k\cdot \xi} = \left.(\iu \partial_k)^n\,\mathcal{Z}_0(k)\right\vert_{k=\frac{V}{T} \Im\omega^{(1)}}~,
\end{equation}
where we have introduced the generating function $ \mathcal{Z}_0(k) = \int \intd\xi\; \ee^{-\mathcal{S}_{0}(\xi)} \,\ee^{-\iu k\cdot \xi} $. 
Each power of $\iu\partial_k$ evidently produces a factor $\frac{T}{V}$. 
At the same time, performing this derivative explicitly on $\mathcal{Z}_0(k)$, with $\mathcal{S}_{0}(\xi)$ specified in Eq.~(\ref{eqn:esnull}), produces a factor 
\begin{equation}
 \delta = \iu \frac{T}{V}\left[ \omega^{(2)} \right]^{-1}\cdot k = \left[ \omega^{(2)} \right]^{-1}\cdot \omega^{(1)}~,
\end{equation}
which is independent of $\frac{T}{V}$. 

Hence there are two small quantities at hand to establish a perturbative expansion: $\frac{T}{V}$ and $\delta$. 
The smallness of $\frac{T}{V}$ is given here as we are interested in the thermodynamic limit.
The size of $\delta$, however, is controlled by the action itself. 
Whether the expansion in $\delta$ is justified or not depends on the model and must be examined accordingly. 
The explicit calculations presented in the main body of this work shows that in the present version of the PNJL model the expansion in $\delta$ is indeed a good approximation.

We are now in a position to write down the thermal expectation value of a generic function $f$ as an expansion in powers of $\frac{T}{V}$ and $\delta$.
We proceed here with establishing Feynman diagrams for this perturbative approach. 
We write generically
\begin{equation}
\label{eqn:genpartfun}
Z \;=\; \frac1{\mathcal{N}}\int \mathcal{D}\xi\;\ee^{-\mathcal{S}_{\mathrm{bos}}} \;=\; \frac1{\mathcal{N}}\int \mathcal{D}\xi\;\;\sum_{l=0}^{\infty}\frac1{l!}\,\left(-\mathcal{S}_{\mathrm{I}}\right)^{l}\;\,\ee^{-\mathcal{S}_0}\,.
\end{equation}
If corrections to the partition function of the PNJL model are to be calculated, the $\mathcal{S}_0$ part of the action only comprises zeroth and second order terms, while the ``small'' part $\mathcal{S}_\mathrm{I}$ is identified with all other orders. 
The first order term acts as a source term.  
We establish the following Feynman rules:
\begin{align}
\label{eqn:feynrules}
\parbox{40pt}{\begin{picture}(40,40)
\put(0,20){\includegraphics{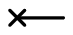}}
\put(10,10){$j$}
\end{picture}} &= - \frac{\partial \mathcal{S}_{\mathrm{bos}}}{\partial \xi_j} &
\parbox{40pt}{\begin{picture}(40,40)
\put(0,20){\includegraphics{./graph2.eps}}
\put(2,10){$j$}\put(32,10){$k$}
\end{picture}} &= +\left[  \frac{\partial^2 \mathcal{S}_{\mathrm{bos}}}{\partial \xi_j \partial \xi_k}\right]^{-1} \nonumber \\
\parbox{40pt}{\begin{picture}(40,40)
\put(0,2){\includegraphics{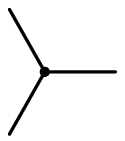}}
\put(6,2){$j$}\put(6,34){$k$}\put(27,10){$l$}
\end{picture}} &=  - \frac{\partial^3 \mathcal{S}_{\mathrm{bos}}}{\partial \xi_j \partial \xi_k \partial \xi_l} &
\parbox{40pt}{\begin{picture}(40,40)
\put(0,0){\includegraphics{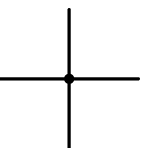}}
\put(2,10){$j$}\put(22,0){$k$}\put(14,32){$l$}\put(30,23){$m$}
\end{picture}} &= - \frac{\partial^4 \mathcal{S}_{\mathrm{bos}}}{\partial \xi_j \partial \xi_k \partial \xi_l \partial \xi_m} \\
\vdots & &\vdots& \nonumber
\end{align}

In perturbation theory it can be shown that only connected diagrams contribute to the partition function, i.\,e.\ 
\begin{equation}
\label{eqn:pertpartition}
Z_\mathrm{I} = \braket{\ee^{-\mathcal{S}_\mathrm{I}}}_0 = \sum_{l=0}^{\infty}\frac1{l!}\braket{(-\mathcal{S}_\mathrm{I})^l}_0 = \exp\left\lbrace \sum_{n=1}^{\infty} \frac1{n!} \braket{(-\mathcal{S}_\mathrm{I})^n}_{0c}\right\rbrace,
\end{equation}
where $\braket{\cdots}_0$ denotes the expectation value with respect to the unperturbed action, and $\braket{\cdots}_{0c}$ is the expectation value of the connected diagrams with respect to this unperturbed action.
Note that here the corrections depicted by the Feynman diagrams are corrections to the negative action, $-\mathcal{S}$, as the partition function was defined by $Z=\ee^{-\mathcal{S}_{\mathrm{eff.}}}$. The corrections therefore need to be subtracted from the mean field result of the action $\mathcal{S}_{\mathrm{MF}}$.

For the thermal expectation values of $f$ we write 
\begin{equation}
\label{eqn:fplusinteraction}
\braket{f} = \braket{f\,\ee^{-\mathcal{S}_{\mathrm{I}}} }_0 = \sum_{l=0}^{\infty} \frac1{l!}\,\braket{f\,(-\mathcal{S}_{\mathrm{I}})^l}_0 .
\end{equation}
Here each term under the sum can be written in terms of connected expectation values
\begin{multline}
\braket{f\,(-\mathcal{S}_{\mathrm{I}})^l}_0 = \sum_{a_1, a_2\cdots,\,a_n,\,m = 0}^{\infty} \frac{l!}{a_1!a_2!(2!)^{a_2}\cdots (a_n!)(n!)^{a_n}m!} \braket{(-\mathcal{S}_{\mathrm{I}})}^{a_1}_{0c} \braket{(-\mathcal{S}_{\mathrm{I}})^2}^{a_2}_{0c} \cdots\\
\cdots \braket{f\,(-\mathcal{S}_{\mathrm{I}})^{m}}_{0c} \delta_{\nu,\,l}~,
\end{multline}
where $\nu = a_1+2a_2+\cdots+na_n+m$. 
Substituting back in Eq.~(\ref{eqn:fplusinteraction}) gives
\begin{equation}
\braket{f\,e^{-\mathcal{S}_{\mathrm{I}}} }_0 = \exp\left\lbrace \sum_{n=1}^{\infty} \frac1{n!} \braket{(-\mathcal{S}_\mathrm{I})^n}_{0c}\right\rbrace\;\times\; \sum_{m=0}^{\infty} \frac1{m!} \braket{f\,(-\mathcal{S}_\mathrm{I})^m}_{0c}.
\end{equation}
Using Eq.~(\ref{eqn:pertpartition}) we find the final result
\begin{equation}
\label{eqn:expvaluef}
\braket{f} = \braket{f\,e^{-\mathcal{S}_{\mathrm{I}}} }_0 = \sum_{n=0}^{\infty} \frac1{n!} \braket{f\,(-\mathcal{S}_\mathrm{I})^n}_{0c}.
\end{equation}
In terms of Feynman diagrams Eq.~(\ref{eqn:expvaluef}) can be translated into all those connected diagrams that contain exactly one insertion coming from the function $f$. The Feynman rules for the insertions of $f$ are
\begin{align}
\label{eqn:feynrulesforf}
\parbox{40pt}{\begin{picture}(40,40)
\put(0,20){\includegraphics{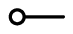}}
\put(10,10){$j$}
\end{picture}} &= \frac{\partial f}{\partial \xi_j} &
\parbox{40pt}{\begin{picture}(40,40)
\put(0,20){\includegraphics{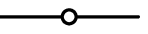}}
\put(2,10){$j$}\put(32,10){$k$}
\end{picture}} &= \frac{\partial^2 f}{\partial \xi_j \partial \xi_k} \nonumber \\
\parbox{40pt}{\begin{picture}(40,40)
\put(0,2){\includegraphics{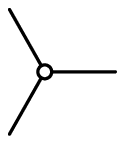}}
\put(6,2){$j$}\put(6,34){$k$}\put(27,10){$l$}
\end{picture}} &=  \frac{\partial^3 f}{\partial \xi_j \partial \xi_k \partial \xi_l} &
\parbox{40pt}{\begin{picture}(40,40)
\put(0,0){\includegraphics{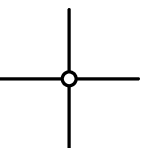}}
\put(2,10){$j$}\put(22,0){$k$}\put(14,32){$l$}\put(30,23){$m$}
\end{picture}} &=  \frac{\partial^4 f}{\partial \xi_j \partial \xi_k \partial \xi_l \partial \xi_m} \\
\vdots & &\vdots& \nonumber
\end{align}
What is needed to use these rules systematically is a scheme that orders all possible diagrams according to their importance in powers of the small parameters $\frac{T}{V}$ and $\delta$.
The lowest order corrections in $\frac{T}{V}$ and $\delta$ are shown in Table~\ref{tab:feynmangraphs}.

\begin{table}
\begin{center}
\begin{tabular}{c||c|c|c||}
\rule[-6pt]{0pt}{20pt} & $\beta = 0$ & $\beta = 1$ & $\beta = 2$ \\ \hline \hline
$\alpha = 0$\rule[-17pt]{0pt}{40pt} & 
$f(\theta_{\mathrm{MF}})$ & 
\parbox{15pt}{\includegraphics{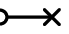}} & \parbox{139pt}{$\frac12\,\,$\parbox{55pt}{\includegraphics{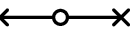}}
$\frac12\,\,$\parbox{55pt}{\includegraphics{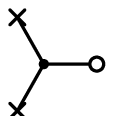}} }    \\ \hline
$\alpha = 1$\rule[-61pt]{0pt}{130pt} & 
\parbox{65pt}{$\frac12\,\,$\parbox{15pt}{\includegraphics{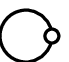}} \\[10pt] $\frac12\,\;$\parbox{45pt}{\includegraphics{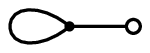}} } & 
\parbox{65pt}{$\frac12\,\,$\parbox{40pt}{\includegraphics{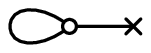}}\\[10pt]
$\frac12\,\,$\raisebox{0pt}{\includegraphics{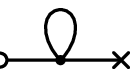}}\\[10pt]
$\frac12\,\,$\parbox{40pt}{\includegraphics{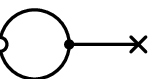}} } & 
\parbox{139pt}{\parbox{65pt}{
$\frac14\,\,$\raisebox{0pt}{\includegraphics{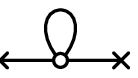}} \\[10pt]
$\frac12\,\,$\raisebox{0pt}{\includegraphics{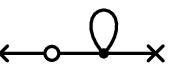}} \\[10pt]
$\frac12\,\,$\parbox{45pt}{\includegraphics{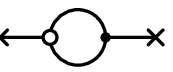}} \\[10pt]
$\frac12\,\,$\parbox{30pt}{\includegraphics{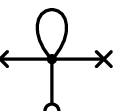}}  }
\parbox{65pt}{
$\frac14\,\,$\parbox{35pt}{\includegraphics{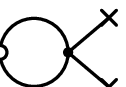}} \\[10pt]
$\frac14\,\,$\parbox{55pt}{\includegraphics{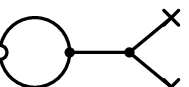}} \\[10pt]
$\frac14\,\,$\parbox{55pt}{\includegraphics{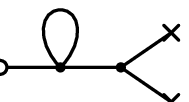}} \\[10pt]
$\frac14\,\,$\raisebox{0pt}{\includegraphics{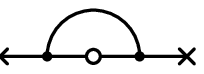}}} } \\ \hline 
$\alpha = 2$ \rule[-115pt]{0pt}{235pt} &  \parbox{65pt}{
$\frac18\,\;$\parbox{40pt}{\includegraphics{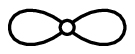}}  \\[10pt]
$\frac{1}{3!}\,\;$\parbox{40pt}{\includegraphics{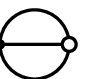}}  \\[10pt]  
$\frac14\,\;$\parbox{40pt}{\includegraphics{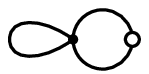}}  \\[10pt]  
$\frac14\,\;$\parbox{40pt}{\includegraphics{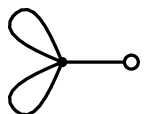}}  \\[10pt]  
$\frac{1}{3!}\,\;$\parbox{40pt}{\includegraphics{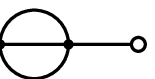}}  } & 
\parbox{139pt}{\parbox{65pt}{
$\frac1{3!}\,\,$\parbox{40pt}{\includegraphics{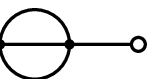}} \\[10pt] 
$\frac18\,\,$\parbox{40pt}{\includegraphics{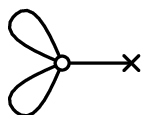}} \\[10pt] 
$\frac1{3!}\,\,$\parbox{40pt}{\includegraphics{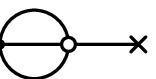}} \\[10pt] 
$\frac1{3!}\,\,$\parbox{40pt}{\includegraphics{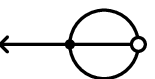}} \\[10pt] 
$\frac14\,\,$\parbox{55pt}{\includegraphics{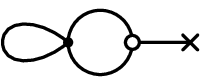}} \\[10pt] $\frac14\,\,$\parbox{40pt}{\includegraphics{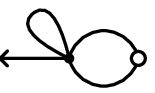}} \\[10pt]
$\frac18\,\,$\parbox{40pt}{\includegraphics{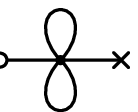}}  }
\parbox{65pt}{
$\frac18\,\,$\parbox{45pt}{\includegraphics{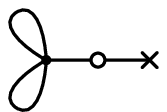}} \\[10pt] 
$\frac1{3!}\,\,$\parbox{50pt}{\includegraphics{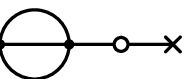}} \\[10pt] 
$\frac1{3!}\,\,$\parbox{40pt}{\includegraphics{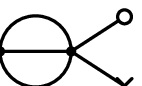}} \\[10pt] 
$\frac1{3!}\,\,$\parbox{50pt}{\includegraphics{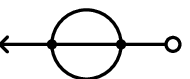}} \\[10pt] 
\parbox{65pt}{\rule{0pt}{0pt}\;\;\includegraphics{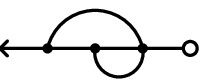}} \\[10pt] 
\parbox{65pt}{\rule{0pt}{0pt}\;\;\includegraphics{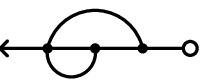}} \\[10pt] 
$\frac14\,\,$\parbox{55pt}{\includegraphics{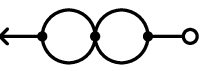}} }} &  \parbox{20pt}{ \vdots  \vspace{40pt}  \vdots  \vspace{40pt} \vdots  } \\ \hline \hline
\end{tabular}
\end{center}\vspace{-12pt}
\caption{\label{tab:feynmangraphs}
The Feynman graphs contributing to $\braket{f}$, ordered in $\left(\frac{T}{V}\right)^\alpha$ and $\delta^\beta$ with multiplicity factors.}
\end{table}

\begin{table}
\begin{center}
\begin{tabular}{c||c|c||}
 & $\beta = 0$ & $\beta = 1$ \\ \hline \hline
$\alpha=0$ \rule[-8pt]{0pt}{20pt}& --- & --- \\  \hline
$\alpha=1$ \rule[-18pt]{0pt}{40pt}& \parbox{65pt}{\centering $\frac12 \times 2$\; \includegraphics{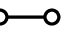}} & 
\parbox{75pt}{\centering $\frac12 \times 4$ \; \includegraphics{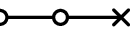}}
\quad $+ \quad\frac12 \times 2$\parbox{45pt}{\centering \includegraphics{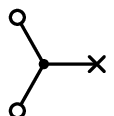}}
\\ \hline \hline
\end{tabular}
\end{center}\vspace{-12pt}
\caption{\label{tab:chisqr} The lowest order Feynman graphs contributing to $\chi_g^2$. The vertices depicted as a circle are now the contributions of $g$, defined analogously to the contributions of $f$ in Eq.~(\ref{eqn:feynrulesforf}). The prefactors are the product of the multiplicity factors of the original Feynman graphs and the factors arising from the differentiation.}
\end{table}

A useful consistency check is to verify that the thermal expectation values are now closer to the properties of an order parameter than the mean field result.
In other words: we examine whether the thermodynamic potential $\Omega$ is a Landau effective action minimised with respect to $\braket{\sigma},\, \braket{\Delta},\, \braket{\Phi},\,\braket{\Phi^*}$ using Eq.~(\ref{eqn:expvaluef}) for the expectation values.
The analysis below is done for the lowest order terms, $\alpha = 0$ and $\beta = 0,\,1$.
We start from the form also used for the numerical calculations, presented below Eq.~(\ref{eqn:numOmega}), and differentiate with respect to the expectation values $\braket{\theta} = (\braket{\sigma},\, \braket{\Delta},\, \braket{\Phi},\,\braket{\Phi^*})^T$. To orders $\alpha = 0$ and $\beta = 0,\,1$ we find that $\braket{\theta} = \theta_0+\delta \theta$, where $\delta \theta$ is given 
by 
\begin{equation}
\delta \theta_i \;=\; \frac12\Bigg( \left. \left[ \frac{\partial^2 \Omega_{0}}{\partial \theta^2} \right]^{-1} \cdot \frac{\partial \Omega_{0}}{\partial \theta}  \right\vert_{\theta= \theta_{\mathrm{MF}}} \Bigg)_i
\label{eqn:fieldshift}
\end{equation}
(which is Eq.~(\ref{eqn:numfield}) with $f(\theta) = \theta_i$ ).
After some calculation we arrive at the lowest order term in $\beta$ 
\begin{multline}
\left.\frac{\partial \Omega}{\partial \theta_i} \right\vert_{\theta = \braket{\theta}} \;=\; 
\frac{9}{8} \sum_{jk}\left[\frac{\partial^3 \Omega_{0}}{\partial \theta^3} \right]_{ijk} \;\;\Bigg(
\left[ \frac{\partial^2 \Omega_{0}}{\partial \theta^2} \right]^{-1} \cdot \frac{\partial \Omega_{0}}{\partial \theta} \Bigg)_j \\
\left.\Bigg(\left[ \frac{\partial^2 \Omega_{0}}{\partial \theta^2} \right]^{-1} \cdot  \frac{\partial \Omega_{0}}{\partial \theta} \Bigg)_k \;\;
\right\vert_{\theta= \theta_{\mathrm{MF}}}\;\cdots \;+\; \text{higher orders},
\end{multline}
which is of order $\beta=2$, i.\,e.\ the self consistency equations are satisfied to the order we have been working in. 
As a consequence the corrections necessary to account for the fermion sign problem do not modify the mean field equations. 
A backward reaction on the mean field equations does not occur at this level of the approximation, which focuses on the evaluation of corrections to the effective (Polyakov loop) potential.
Additional effects of pionic and scalar quark-antiquark modes, as considered in Sec.~\ref{sec:pions}, do in principle have backward effects on the mean field equation. 

The formalism allows to determine susceptibilities involving a quantity $g$, $
\chi_g = [ V (\braket{g^2}-\braket{g}^2 ) ]^{1/2}$.
All that needs to be done is to apply the previously developed formalism to the function $g^2$.
In Table~\ref{tab:feynmangraphs} the Feynman rules and multiplicity factors are written down for the evaluation of $\braket{f}$. In a second step $f$ is replaced by $g^2$. In this step the product rule of differentiation has to be applied producing additional prefactors. 
In this procedure it will happen that vertices of $f$ with $m=2,\, 3,\dots$ or more legs will split into two vertices with $m_1+m_2 =m$ legs. % according to Eqs.~(\ref{eqn:g2derivs}). 
The lowest orders of the expression, Eq.~(\ref{eqn:genericsusc}), are shown in Table~\ref{tab:chisqr}.
The contributions of order $\left(\frac{T}{V}\right)^0$ cancel.
In this framework susceptibilities scale with $V^{\frac12}$ as expected.
Additionally, it becomes obvious from Table~\ref{tab:chisqr} that there are no mean field contributions to susceptibilities in the sense that $\braket{(g-\braket{g}_{\mathrm{MF}})^2}_{\mathrm{MF}} = \braket{g^2}_{\mathrm{MF}}-\braket{g}^2_{\mathrm{MF}} = g^2_{\mathrm{MF}}-g^2_{\mathrm{MF}} = 0$. 
In the framework of mean field calculations, susceptibilities are usually evaluated by inverting the second derivative of the mean field action with respect to the fields. 
This is seen in the present framework as well: the entry for $\alpha=1$ and $\beta=0$ in Table~\ref{tab:chisqr} produces exactly this expression.

\end{document}